\documentclass[iop,apj]{emulateapj}
\usepackage{graphicx}

\newcommand{\grav}{log($g$)}
\newcommand{\teff}{$T_{\mbox{\scriptsize eff}}$}
\newcommand{\kms}{km~s$^{-1}$}

\newcommand{\subsun}{\mbox{$_{\odot}$}}

\shorttitle{Bizarre Abundances in NGC 2419}
\shortauthors{Cohen \& Kirby}
\slugcomment{Accepted to ApJ, 2012 September 12}

\begin{document}

\title{The Bizarre Chemical Inventory of NGC~2419, An Extreme Outer Halo Globular Cluster\altaffilmark{1}}

\author{Judith G. Cohen\altaffilmark{2}
and Evan N. Kirby\altaffilmark{2,3}  }

\altaffiltext{1}{Based in part on observations obtained at the
W.M. Keck Observatory, which is operated jointly by the California
Institute of Technology, the University of California, and the
National Aeronautics and Space Administration.}

\altaffiltext{2}{Palomar Observatory, Mail Stop 249-17,
California Institute of Technology, Pasadena, Ca., 91125,
jlc(enk)@astro.caltech.edu}

\altaffiltext{3}{Hubble Fellow}

\begin{abstract}

We present new Keck/HIRES observations of six red giants in the
globular cluster NGC~2419.  Although the cluster is among the most
distant and most luminous in the Milky Way, it was considered
chemically ordinary until very recently.  Our previous work showed
that the near-infrared \ion{Ca}{2} triplet line strength varied more
than expected for a chemically homogeneous cluster, and that at least
one star had unusual abundances of Mg and K\@.  Here, we confirm that
NGC~2419 harbors a population of stars, comprising about one third of
its mass, that is depleted in Mg by a factor of 8 and enhanced in K by
a factor of 6 with respect to the Mg-normal population.  Although the
majority, Mg-normal population appears to have a chemical abundance
pattern indistinguishable from ordinary, inner halo globular clusters,
the Mg-poor population exhibits dispersions of several elements.  The
abundances of K and Sc are strongly anti-correlated with Mg, and some
other elements (Si and Ca among others) are weakly anti-correlated with Mg.
These abundance patterns suggest that the different populations of
NGC~2419 sample the ejecta of diverse supernovae in addition to AGB
ejecta.  However, the abundances of Fe-peak elements except Sc show no
star-to-star variation.  We find no nucleosynthetic source that
satisfactorily explains all of the abundance variations in this
cluster.  Because NGC~2419 appears like no other globular cluster, we
reiterate our previous suggestion that it is not a globular cluster at
all, but rather the core of an accreted dwarf galaxy.

\end{abstract}


\section{Introduction \label{section_intro} }

NGC~2419 is one of the most unusual globular clusters (GCs) belonging
to the Milky Way (MW)\@.  It resides in the MW's outer halo
\citep[][90~kpc from the Galactic center]{harris97}.  It is notable
not just for its distance but also its luminosity.  M54, the core of
the Sagittarius dwarf galaxy \citep{ibata95}, is the only GC more
luminous than NGC~2419.

The first {\it Hubble Space Telescope} (HST) photometry
\citep{harris97} of NGC~2419 showed that the cluster is as old as the
inner halo cluster M92\@.  In other words, NGC~2419 is about as old as
the Universe.  Like many GCs, NGC~2419 is a ``second parameter''
cluster, with an extended blue horizontal branch (HB)\@.
\citet*{dicriscienzo11} attributed the HB morphology and the color
dispersion at the base of the red giant branch (RGB) to a different
helium abundance between the first and second generations of stars, a
popular explanation for the second parameter in GCs \citep{dantona02}.

The distance and photometric properties of the cluster alone are not
extremely unusual, but the chemical properties of the cluster are.
Early spectroscopy \citep{suntzeff88} did not reveal any unusual
abundance patterns in the cluster.  In particular, the iron abundance
appeared invariable from star to star.  Much more recently,
medium-resolution (Keck/DEIMOS) spectroscopy by \citet[][hereafter
  C10]{deimos} showed that the strength of the infrared calcium
triplet (CaT) varies from star to star, even at fixed stellar
luminosity.  \citet{ibata11} confirmed this star-to-star variation
with independent DEIMOS spectroscopy.
\citeauthor*{deimos}\ attributed this variation to a range of calcium
abundance\footnote{The 
standard nomenclature is adopted; the abundance of
element $X$ is given by $\epsilon(X) = N(X)/N(H)$ on a scale where
$N(H) = 10^{12}$ H atoms.  Then
[X/H] = log$_{10}$[N(X)/N(H)] $-$ log$_{10}$[N(X)/N(H)]\subsun, and similarly
for [X/Fe].}
($\sigma({\rm [Ca/H]}) \sim 0.2$) in the cluster.

Later high-resolution (Keck/HIRES) spectroscopy by \citet[][hereafter
  C11]{cohen11} revealed an even more complex abundance distribution.
Six of seven stars appeared identical to stars in ``normal'' GCs, such
as those found in the inner halo of the MW\@.  However, star S1131 had
${\rm [K/Fe]} = 1.1$ (very high for a GC) and ${\rm [Mg/Fe]} = -0.5$
(very low for a GC)\@.  Such a low value of [Mg/Fe] can be found only
in the most metal-rich stars in dwarf spheroidal galaxies
\citep{letarte10,cohen_umi,kirby11}.  It is never found in stars with the
metallicity of NGC~2419 (${\rm [Fe/H]} = -2.1$).  Furthermore, low
values of [Mg/Fe] in halo and dwarf galaxy stars are always
accompanied by low abundances of other $\alpha$ elements, such as Si
and Ca.  Star S1131 exhibits enhanced ratios of [Si/Fe] and [Ca/Fe],
typical for a normal GC star.

\citeauthor*{cohen11}'s HIRES sample provided no compelling evidence
for a variation in elements heavier than potassium, such as Fe and
other iron-peak elements.  Even Ca, in which a dispersion was detected
with DEIMOS, appeared to be constant across the stars from HIRES data.
Importantly, the DEIMOS analysis was based on ionized Ca lines whereas
the HIRES analysis was based on neutral Ca lines.

The HIRES sample of \citeauthor*{cohen11} included only one of the
Ca-rich stars identified by \citeauthor*{deimos}\@.  In fact, that
single star was S1131, the one with unusual magnesium and potassium
abundances.  The peculiarity of this star with a strong CaT demanded
that we observe additional stars from the DEIMOS sample with large CaT
line strengths.  In this article, we expand \citeauthor*{cohen11}'s
sample of HIRES spectra in NGC~2419, focusing in particular on the
stars with strong CaT lines.

\section{Observations and Abundance Analysis Procedures 
\label{section_obs_anal} }

In an effort to overcome some of the limitations of and concerns
arising from our previous work in NGC 2419, we obtained HIRES-R
\citep{vogt_hires} spectra of an additional 6 stars in this GC during
a 4 night run which began 2012 Jan 29.  The spectrograph configuration
was identical to that we used in our 2008 and 2010 observations of NGC
2419 red giants.  Most of the new stars were chosen to probe the
Ca-rich distribution of the DEIMOS measurements of the Ca triplet,
with preference given to those stars for which we had already obtained
low signal-to-noise ratio (S/N) HIRES spectra that suggested, 
based on our measured radial
velocities, that the stars were cluster members.  These stars are
faint ($17.5 < V < 17.9$) for high-resolution spectroscopy, and the
exposure times ranged from 2.5 hours for the brightest star to 4 hours
for the faintest star.  All of the nights were clear, and two had
excellent seeing.

\begin{deluxetable*}{l l rrr rr}
\tablewidth{0pt}
\tablecaption{Data for Red Giant Members of NGC~2419 With 
Keck/HIRES Spectra
\label{table_sample} }
\tablehead{
\colhead{Name\tablenotemark{a}} & \colhead{V\tablenotemark{a}} &
  \colhead{$T_{\rm eff}$, log($g$),$v_t$\tablenotemark{b}} & 
  \colhead{Exp. Time} & \colhead{S/N\tablenotemark{c}} & 
  \colhead{$v_r$} \\
\colhead{} & \colhead{(mag)} & \colhead{(K,~dex,~km s$^{-1}$)} &
   \colhead{(sec)} &
  \colhead{ } & \colhead{(km s$^{-1}$)}
}
\startdata
New Stars \\
Stet~406 & 17.80 & 4448, 0.95 & 14400 & 95 & $-20.3$ \\
Stet~458 & 17.90 & 4480, 1.02 & 10800 & 85 & $-17.9$ \\
Stet~1004 & 17.91 & 4482, 1.02 & 14400 & 80 &  $-23.5$ \\
Stet~1065 & 17.66 & 4455, 0.88 & 10800 & 80 & $-22.7$ \\
Stet~1166 & 17.50 & 4350, 0.89 &  9000 & 82 & $-21.3$ \\
Stet~1673 & 17.68 & 4409, 0.88 & 10800 & 95 & $-24.6$ \\
~ \\
From C11 \\ 
Stet~223\tablenotemark{d} &
         17.25 & 4265, 0.61 & 8400 & $>100$ & $-22.4$ \\
Stet~810 & 17.31 & 4316, 0.65 &   7200 & $>100$ & $-22.6$ \\
Stet~973\tablenotemark{e} &
      17.45 & 4325, 0.74 &  3200 & 39 & $-21.9$\\
Stet~1131 & 17.61  & 4382, 0.84 &  9000 & 95 & $-16.8$  \\
Stet~1209\tablenotemark{f} &
         17.41 & 4317, 0.71 & 7200 & 93 & $-19.0$ \\
Stet~1305\tablenotemark{g} & 
      17.61 & 4385, 0.84 &  3000 & 52 & $-16.8$ \\
Stet~1814\tablenotemark{h} & 
       17.27 &  4472, 0.62 & 5400 & 90 & $-26.1$ \\
\enddata
\tablenotetext{a}{Star IDs and $V$~magnitudes are from the
online version of the database of \cite{stetson05}.}
\tablenotetext{b}{These values, adopted for the
abundance analysis, are based only on the $V$ mag and an appropriate isochrone.}
\tablenotetext{c}{SNR per spectral resolution element
in the continuum at 5500~\AA.}
\tablenotetext{d}{Stetson~223 = Suntzeff~1.}
\tablenotetext{e}{Stetson~973 = Suntzeff~15.}
\tablenotetext{f}{Stetson~1209 = Suntzeff~16.}
\tablenotetext{g}{Stetson~1305 = RH~10 
  (private communication from M.~Shetrone),
  previously observed by \cite{shetrone01}.}
\tablenotetext{h}{Stetson~1814 = Suntzeff~14.}
\end{deluxetable*}

\begin{figure}
\centering
\includegraphics[width=\linewidth]{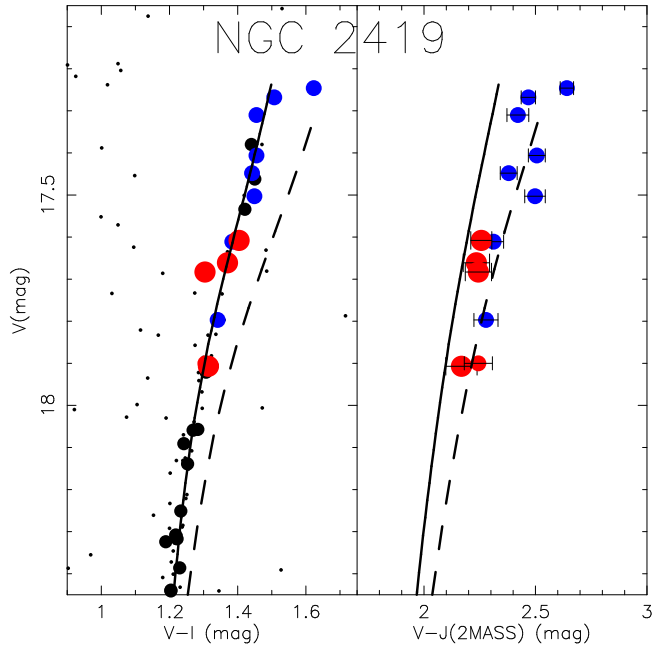}
\caption[]{The $V,~V-I$ (left panel) and $V,~V-J$ (right panel) CMDs
  are shown for NGC~2419 with optical photometry from \cite{stetson05}
  and $J$ photometry from 2MASS.  The red circles denote stars with
  [Mg/Fe] $< 0$, while the blue circles represent stars with [Mg/Fe]
  $> 0$~dex.  The larger symbols denote stars whose DEIMOS spectra
  have near-IR Ca triplet lines implying ${\rm [Ca/H]} > -1.8$.  Two
  12~Gyr, $\alpha$-enhanced isochrones from \cite{yi03} are shown:
  ${\rm [Fe/H]} = -2.2$ (solid line) and ${\rm [Fe/H]} = -1.9$ (dashed
  line).  The smallest black points, seen only in the left panel, are
  the photometric sample of \cite{stetson05}.  The somewhat larger
  black points (also shown only in the left panel) are the Keck/DEIMOS
  sample of C10.
\label{figure_cmd}}
\end{figure}

The total sample of 13 stars for which we obtained reasonably good
HIRES spectra is given in Table~\ref{table_sample}.
Fig.~\ref{figure_cmd} shows a comparison of the location of the sample
stars on the observed $V$, $V-I$ plane using optical photometry from
\cite{stetson05} and the $V$, $V-J$ plane, where $J$ is from 2MASS
\citep{2mass1,2mass2}.  Isochrones from the Y$^2$ grid \citep{yi03}
for an age of 12~Gyr with [Fe/H] = $-2.20$~dex (solid line) and [Fe/H]
= $-1.90$~dex (dotted line), both with [$\alpha$/Fe] = +0.30~dex, are
indicated.  We adopted the same distance and reddening as was used in
\citeauthor*{cohen11}.  In our earlier work (\citeauthor*{deimos}),
the membership of star S1673 in NGC~2419 was considered possible but
not definite.  This is because, as is seen in Fig~\ref{figure_cmd},
this star lies to the blue of the main cluster RGB in a $V$, $V-I$
CMD\@.  Although its $v_r$ is consistent with cluster membership, we
decided to be cautious and not consider it a confirmed member at that
time.  However, on the basis of its abundances, discussed below, we
now consider S1673 to be a definite member of this GC\@.  It may be, as
\citeauthor*{deimos} speculated, an AGB rather than a RGB star.

\begin{figure}
\centering
\includegraphics[width=\linewidth]{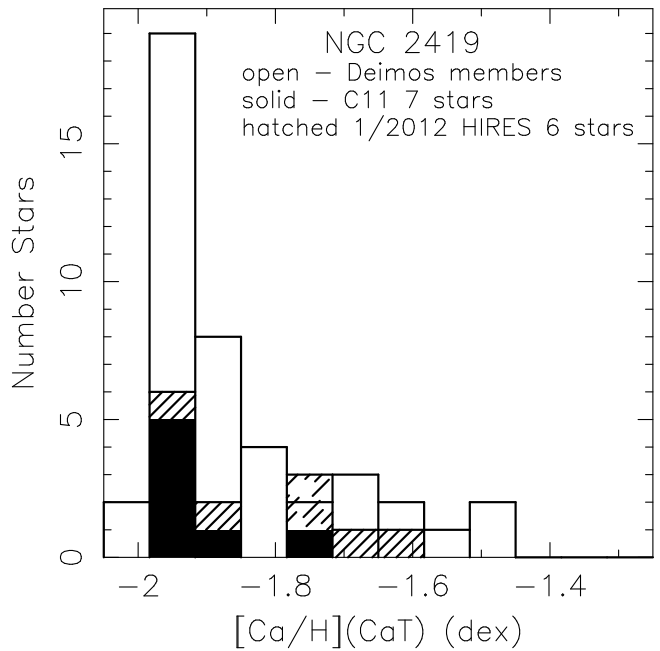}
\caption[]{A histogram of [Ca/H] as inferred from the near-IR
  \ion{Ca}{2} triplet line strengths in the DEIMOS moderate resolution
  spectra of \citeauthor*{deimos} is shown for the sample of 43
  definite members of NGC~2419 isolated in that paper.  The sample
  from C11 of 7 RGB stars in this GC with HIRES spectra is shown by
  the solid fill.  The 6 additional NGC~2419 stars presented here are
  indicated by the hatched areas.
\label{figure_cah_hist}}
\end{figure}

Fig.~\ref{figure_cah_hist} shows a histogram of Ca(CaT), the Ca
abundance inferred from our initial moderate resolution study
(\citeauthor*{deimos}) based on the CaT line strength measured with
DEIMOS on Keck~II, with the HIRES sample indicated.  Our HIRES sample
now includes giants spanning almost the entire range of Ca(CaT)
abundances for NGC~2419 luminous giants.  While there are a few stars
in our DEIMOS sample with even higher Ca(CaT), they are not in our
HIRES sample as they are all fainter than $V = 17.7$.

The measurement of equivalent widths, whose values are given in
Tables~\ref{table_eqw_page1} and \ref{table_eqw_page2}, and the
abundance analyses were carried out in a manner identical to our
previous work.  \citeauthor*{cohen11} described those procedures in
detail.  $W_{\lambda}$ for the 7 stars from \citeauthor*{cohen11} are
also listed there as a number of lines were added since 2011.
Hyperfine structure corrections have been made following
\citeauthor*{cohen11}.  No non-LTE corrections were made because the
Al abundances were calculated not from the 3950~\AA\ resonance doublet
but from the weak 6696,~6698~\AA\ doublet, which has no strong non-LTE
correction \citep[see][]{al_nonlte}.  The magnitudes of the expected
non-LTE corrections for some other key elements are discussed in
\S\ref{section_abund_spread}.  Because these are rather faint
metal-poor giants, we included the strong Mg triplet lines and
sometimes the Na~D lines in the analysis in order to get a reasonable
number of lines for these key elements.  We provide two measurements
of abundance for these two elements: one with and one without these
very strong lines.

We made two important updates to our procedures described in
\citeauthor*{cohen11}.  We are now using the 2010 version of MOOG
\citep{moog} updated by J.~Sobeck \citep{sobeck_moog}.  The new
version contains a better treatment of coherent, isotropic scattering,
which in the 2002 version is treated as pure absorption.  This could
be important for our NGC~2419 sample primarily because these are cool
luminous RGB stars.  But since [Fe/H] for NGC 2419 is about
$-2.1$~dex, the importance of Rayleigh scattering as an opacity source
is not as large as it would be for even more metal-poor stars.
Furthermore, we did not consider lines blueward of 4100~\AA\ because
the spectral S/N is too low at those wavelengths.  We focused on the
part of the spectra $\lambda > 4500$~\AA, where S/N is higher, unless
an element has no or very few lines beyond 4500~\AA\@.  Thus, for our
sample, the use of the 2010 version of MOOG does not introduce
noticeable changes compared to the 2002 version.

\begin{figure}
\centering
\includegraphics[width=\linewidth]{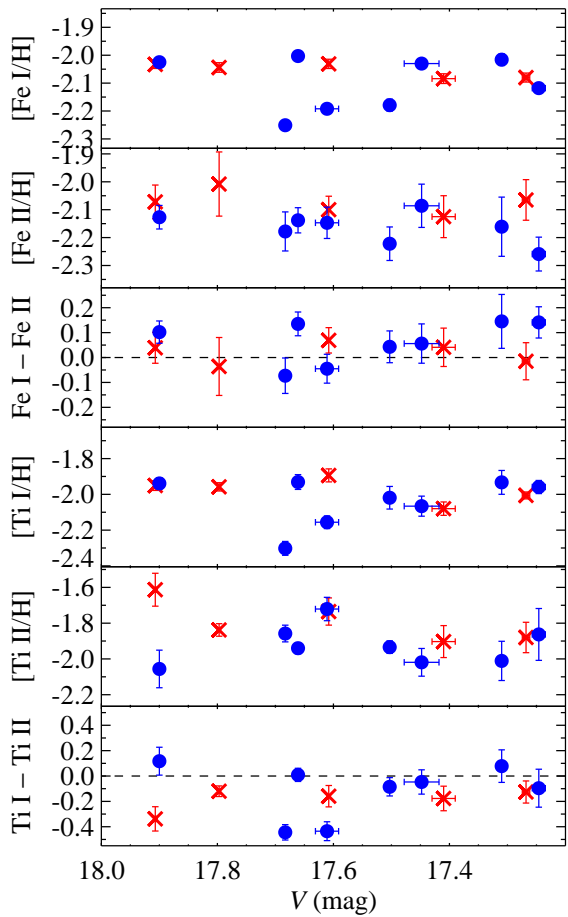}
\caption[]{Diagnostic abundance ratios for two stages of ionization of
  Fe and of Ti are shown as a function of $V$~magnitude for our sample
  of 13 luminous RGB stars in NGC~2419.  Red crosses and blue points
  denote Mg-poor and Mg-rich giants respectively.
\label{figure_anal_diag}}
\end{figure}

The second major change change we made involves the determination of
stellar parameters.  Fig.~\ref{figure_cmd} shows the $V$, $V-I$ and
$V$, $V-J$ CMDs for our HIRES sample in NGC~2419, with two metal-poor,
$\alpha$-enhanced, 12~Gyr Yonsei-Yale isochrones \citep{yi03}
superposed.  We set the stellar parameters \teff\ and \grav\ by
assuming the stars lie on an isochrone halfway between the two shown
in Fig.~\ref{figure_cmd}.  We did not use the colors at all, just the
$V$~magnitude, to set \teff\ and \grav.  When looking for small
abundance variations, the choice of stellar parameters is critical, as
discussed by \citeauthor*{cohen11}.  Rather than relying on colors,
which for such faint stars have non-trivial uncertainties,
particularly those from 2MASS, we decided to force the stars to lie on
an isochrone with just the $V$ measurements taken from
\cite{stetson05}, whose uncertainties are quite small
(${\leq}0.015$~mag), over a total range spanned by our sample of 17.25
to 17.91~mag.  The range in $V-I$ spanned by our sample is only
0.32~mag, less than half of that of $V$ and with somewhat larger
uncertainties.  The range in $V-J$ is 0.47~mag, but the uncertainties
are much larger due to the limited depth of 2MASS\@.  If the stars
really do lie along a single isochrone, as would be the case if
NGC~2419 is actually a chemically homogeneous old GC, using just $V$
will give a very accurate relative \teff\ determination for each star.
Even if a slightly inappropriate isochrone is used, the relative
differences in \teff\ for members along the upper RGB of the stellar
population will be highly accurate.  Fig.~\ref{figure_anal_diag} shows
the dependence of [\ion{Fe}{1}/H], [\ion{Fe}{2}/H] and the difference
of the two as a function of $V$ (our proxy for \teff), as well as the
same for Ti.  The behavior of these key diagnostics serves to
demonstrate that our detailed abundance analyses are valid.

\section{The Chemical Inventory of the NGC~2419 Giants}

Our detailed abundance analysis for 13 luminous red giants in NGC~2419
yielded the results given in Tables~\ref{table_abund_page1} and
\ref{table_abund_page2}.  
The Mg abundances are listed with the two Mg triplet lines  both included
and excluded to illustrate that the result is identical to within
the errors even when the strong Mg triplet lines are included.
The abundance analyses for the 7 stars from
\citeauthor*{cohen11} were redone, resulting only in small
differences.  The present results supersede those of
\citeauthor*{cohen11}.  Tables of uncertainties for the absolute
abundances and the abundance ratios were given in
\citeauthor*{cohen11}.

\setcounter{table}{5}

One star, S1673, appears to be significantly bluer than the RGB of
NGC~2419 in the $V$, $V-I$ CMD, but less discrepant in the $V$, $V-J$
CMD\@.  It may be an AGB star.  We carried out an abundance analysis for
this star based solely on its $V$ magnitude (i.e., assuming that it
lies on the normal RGB) and also one assuming that it is 100~K hotter
than the RGB and with a slightly lower \grav\ corresponding to a mass
of 0.6~$M_{\odot}$ instead of 0.8~$M_{\odot}$ (i.e., on the AGB).  Both sets of results
are presented in Table~\ref{table_abund_page1}.

Our detailed abundance analysis shows that NGC~2419 contains two
groups of stars.  The first, containing 8 of the 13 stars, represents
a normal $\alpha$-rich population typical of GC (and inner halo)
stars.  This is the population that dominated the sample of
\citeauthor*{cohen11}, where it was shown that they are essentially
identical in chemical inventory to the stars in the much nearer, inner
halo cluster NGC~7099 with similar metallicity.

However, the second population found here is very strange.  It shows
extreme depletions of Mg, with [Mg/Fe] ranging widely from $-0.2$ to
$-0.7$~dex, accompanied by large enhancements in K of $\sim$0.7~dex
above those of the ``normal'' NGC~2419 giants.  The Mg-poor group
contains the same 5 stars shown in Fig.~\ref{figure_cah_hist} to have
stronger near-IR Ca triplet lines from the study of
\citeauthor*{deimos}.  
The fraction of stars in the high tail of the Ca(CaT)
distribution (${\rm Ca/H]} > -1.85$) from the larger DEIMOS sample of
  \citeauthor*{deimos} is  34\%; the fraction
of peculiar Mg-poor stars in the smaller HIRES
sample is comparable (38\%).

\begin{deluxetable*}{l  rll rrr}
\tablewidth{0pt}
\tablecaption{Mean Absolute Abundances for the Two Groups of Red Giant Members of NGC~2419 With 
Keck/HIRES Spectra
\label{table_mean_abund} }
\tablehead{
\colhead{ } & \colhead{ } & \colhead{Mg-rich\tablenotemark{a}} &  \colhead{ } & \colhead{ } &
    \colhead{Mg-poor\tablenotemark{b}} \\ 
\colhead{Species} & \colhead{log[$\epsilon$(X)]}   &  \colhead{$\sigma$} & \colhead{N(stars)} &      
    \colhead{log[$\epsilon$(X)]} &  \colhead{$\sigma$} &  \colhead{N(stars)} 
}
\startdata
C(CH)    &    5.75 & 0.19 &  7  &   5.71 & 0.22 & 5 \\
NaI    &   4.31 & 0.29 & 8 &    4.35  & 0.13 & 5 \\
MgI    &    5.78 & 0.14 &   8  &  4.92  & 0.30 &  5  \\
MgI\tablenotemark{c} &        5.74 & 0.14 & 5  & 4.80 & 0.39 & 3 \\
AlI &  4.94 &  0.25 & 4 &  4.93 & 0.11 & 4\tablenotemark{d} \\
SiI &      5.81 & 0.07 & 6 &  6.05 &  0.10 &  5 \\
KI &     3.46 & 0.16 & 8 &   4.26 & 0.13 & 5  \\
CaI\tablenotemark{e}    &  4.40 &  0.09 &  8 & 4.58  & 0.03 & 5 \\
CaT\tablenotemark{f}  & 4.39 & 0.03 & 8 & 4.66 & 0.04 & 5 \\
ScII &  1.10 & 0.04 &  8 &  1.55  & 0.16  & 5 \\
TiI  &      2.95 & 0.13 & 8  & 2.97 &  0.11 & 5 \\
TiII &      3.01 & 0.14 & 8  &  3.12 & 0.12 & 5 \\
TiII\tablenotemark{g} &       3.05 & 0.10 & 8  &  3.21 & 0.12 & 5 \\
VI  &   1.91 &  0.15 &  8&  2.03  & 0.16 &  5  \\
CrI  &   3.21 & 0.14 &  8 &   3.29 & 0.11&  5 \\
MnI  &      2.91 & 0.11 & 8 &   2.94 & 0.12 &  5 \\
FeI  &      5.35 & 0.10 & 8 &   5.37 & 0.10  & 5 \\
FeII &  5.29 & 0.06 &  8  & 5.40 & 0.10 & 5 \\
CoI    &    2.94 & 0.18 & 6 &   3.00 & 0.15 &  5 \\
NiI  &   4.12 & 0.13 & 8 &  4.12 & 0.09 & 5 \\
CuI  &    1.49 & 0.14 & 8 &  1.53 & 0.13 & 5 \\
ZnI  &     2.33 & 0.07 &  5 & 2.47 & 0.09 & 5 \\
YII  &   $-$0.31 & 0.06 & 8 &  $-$0.23 & 0.06 & 5 \\
BaII &    $-$0.12 & 0.12 & 8 &  $-$0.09 & 0.15 & 5  \\ 
CeII  & $-$0.90 & 0.15 & 5 & $-$0.84  & 0.12 & 5 \\
NdII &  $-$0.64 & 0.17 &  7\tablenotemark{d} & $-$0.59 & 0.14 & 4  \\
EuII &   $-$1.44 & 0.15 &  6\tablenotemark{d} & $-$1.32 & 0.34 & 5  \\
\enddata
\tablenotetext{a}{8 RGB stars in NGC~2419 with [Mg/Fe] $> 0$~dex.}
\tablenotetext{b}{5 RGB stars in NGC~2419 with [Mg/Fe] $< 0$~dex.
S1673 is taken as a RGB star; the changes in the means are very small
if it is assumed to be on the AGB.}
\tablenotetext{c}{The two strong Mg triplet lines are excluded.}
\tablenotetext{d}{One upper limit is omitted.}
\tablenotetext{e}{Based on Ca~I lines in the HIRES spectra.}
\tablenotetext{f}{Based on the 8542 and 8662~\AA\ lines in Keck/DEIMOS 
spectra, see C10.}
\tablenotetext{g}{The 4911~\AA\ line of TiII is excluded.}
\end{deluxetable*}

\begin{figure}
\centering
\includegraphics[width=\linewidth]{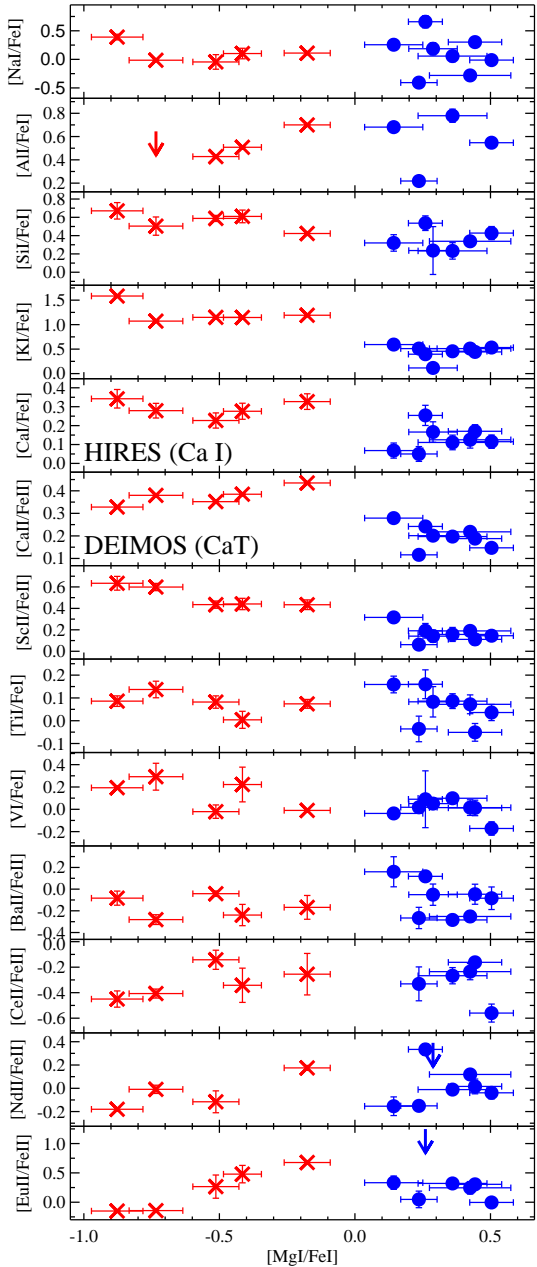}
\caption[]{Abundance ratios for various species with respect to Fe,
  using \ion{Fe}{1} or \ion{Fe}{2} as appropriate, are shown as a
  function of [Mg/Fe] for our sample of 13 giants in NGC~2419 with
  HIRES spectra.  Red crosses and blue points denote Mg-poor and
  Mg-rich giants respectively.
\label{figure_abund_mg}}
\end{figure}

Table~\ref{table_mean_abund} gives the mean abundances for each of the
two groups.  A number of smaller anomalies are apparent from this
table, and are also visible in Fig~\ref{figure_abund_mg}. We see that
the Mg-poor group of luminous RGB stars in NGC~2419 has slightly
higher [Si/Fe], [Sc/Fe], and [Ca/Fe] than does the group of normal
giants.  However the majority of the elements probed have identical
abundances with relatively small dispersions in the two groups of
NGC~2419 stars.  This specifically includes Fe.

\begin{figure}
\centering
\includegraphics[width=\linewidth]{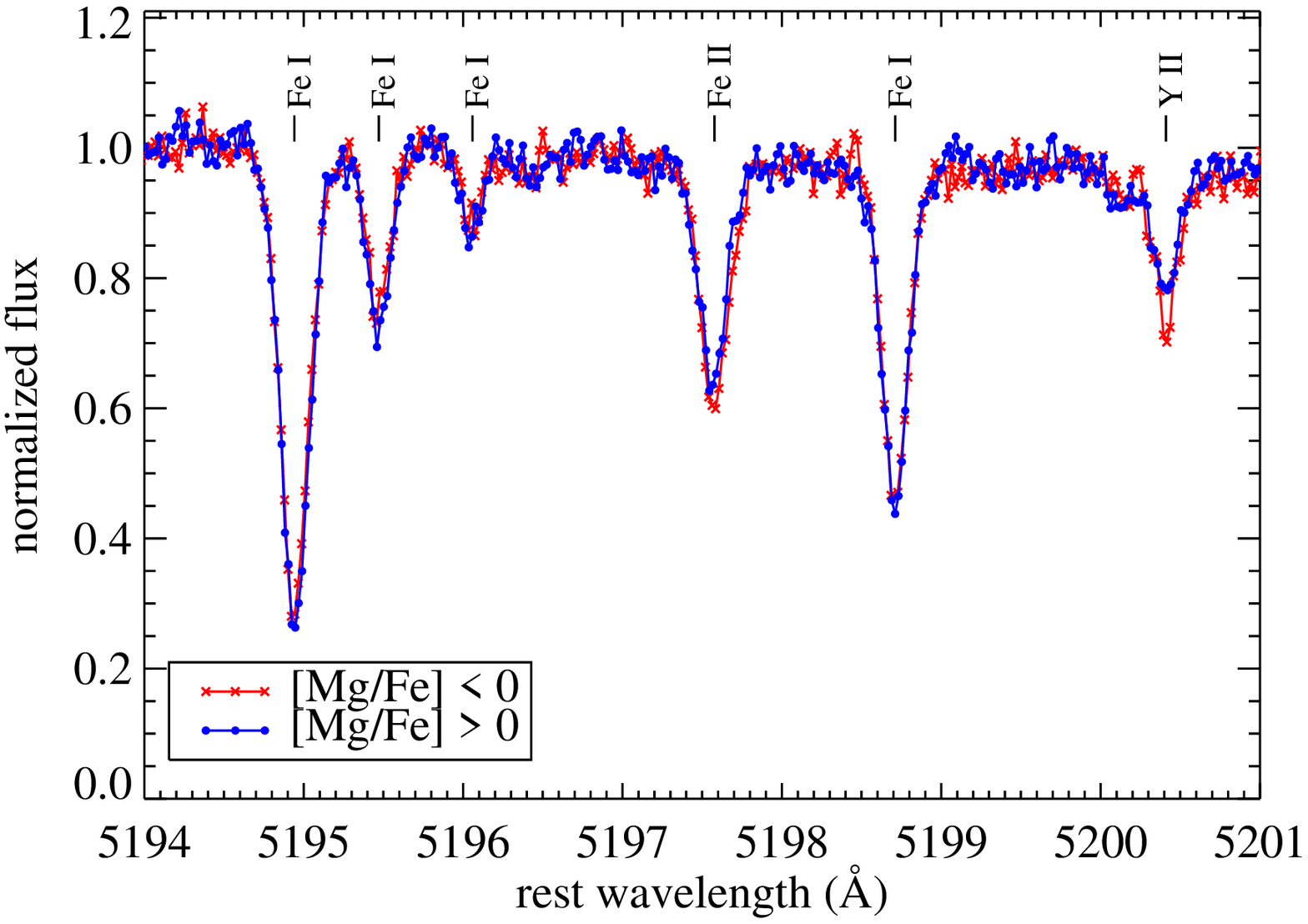}
\caption[]{Sums of sections of spectra for the Mg-normal (blue points)
  and Mg-poor (red crosses) NGC~2419 giants centered on a group of
  \ion{Fe}{1} lines and a \ion{Fe}{2} line at 5197~\AA\@. The two are
  indistinguishable.  The sum for the Mg-normal stars omits S223, the
  most luminous and reddest NGC~2419 giant.
\label{figure_5197}}
\end{figure}
%

\begin{figure}
\centering
\includegraphics[width=\linewidth]{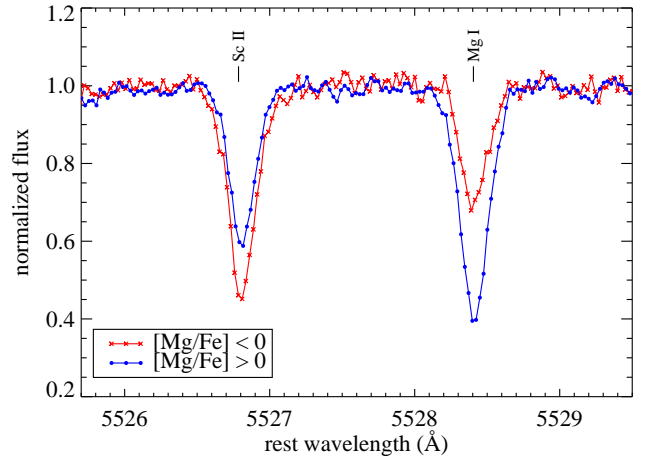}
\caption[]{Sums of sections of spectra for the Mg-normal (blue points)
  and Mg-poor (red crosses) NGC~2419 giants centered on the
  \ion{Mg}{1} line at 5528~\AA\ and the \ion{Sc}{2} line at
  5526~\AA\@.  The extremely large difference in Mg abundance, as well
  as the somewhat smaller difference in Sc abundance, between the two
  groups is apparent. The sum for the Mg-normal stars omits S223, the
  most luminous and reddest NGC~2419 giant.
\label{figure_5528}}
\end{figure}
%

\begin{figure}
\centering
\includegraphics[width=\linewidth]{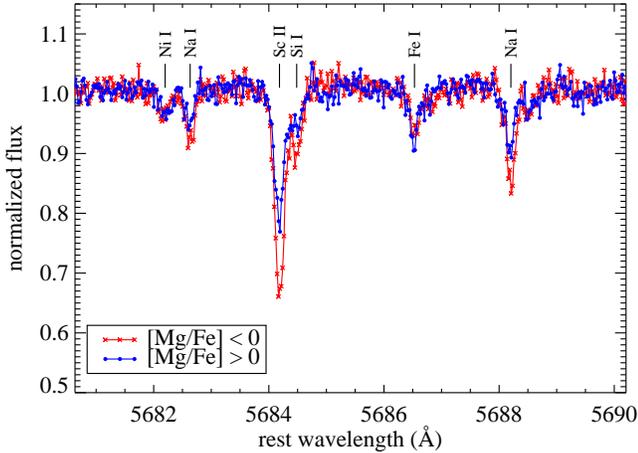}
\caption[]{Sums of sections of spectra for the Mg-normal (red crosses)
  and Mg-poor (blue points) NGC~2419 giants centered on the
  \ion{Na}{1} doublet at 5685~\AA\@.  Note that the \ion{Na}{1} lines
  are only slightly stronger in the Mg-poor stars, while the adjacent
  \ion{Sc}{2} and \ion{Si}{1} lines are significantly stronger in
  those stars.  The sum for the Mg-normal stars omits S223, the most
  luminous and reddest NGC~2419 giant.
\label{figure_5680na}}
\end{figure}

\begin{figure}
\centering
\includegraphics[width=\linewidth]{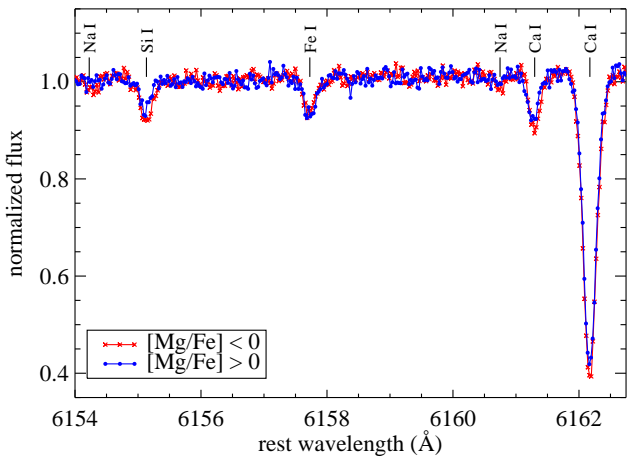}
\caption[]{Sums of sections of spectra for the Mg-normal (red crosses)
  and Mg-poor (blue points) NGC~2419 giants centered on the
  \ion{Na}{1} doublet at 6160~\AA, which is so weak that it is barely
  detectable even in summed spectra.  The adjacent \ion{Ca}{1} and
  \ion{Fe}{1} lines are almost identical between the Mg-poor stars and
  Mg-normal stars.  The sum for the Mg-normal stars omits S223, the
  most luminous and reddest NGC~2419 giant.
\label{figure_6160na}}
\end{figure}
%

\begin{figure}
\centering
\includegraphics[width=\linewidth]{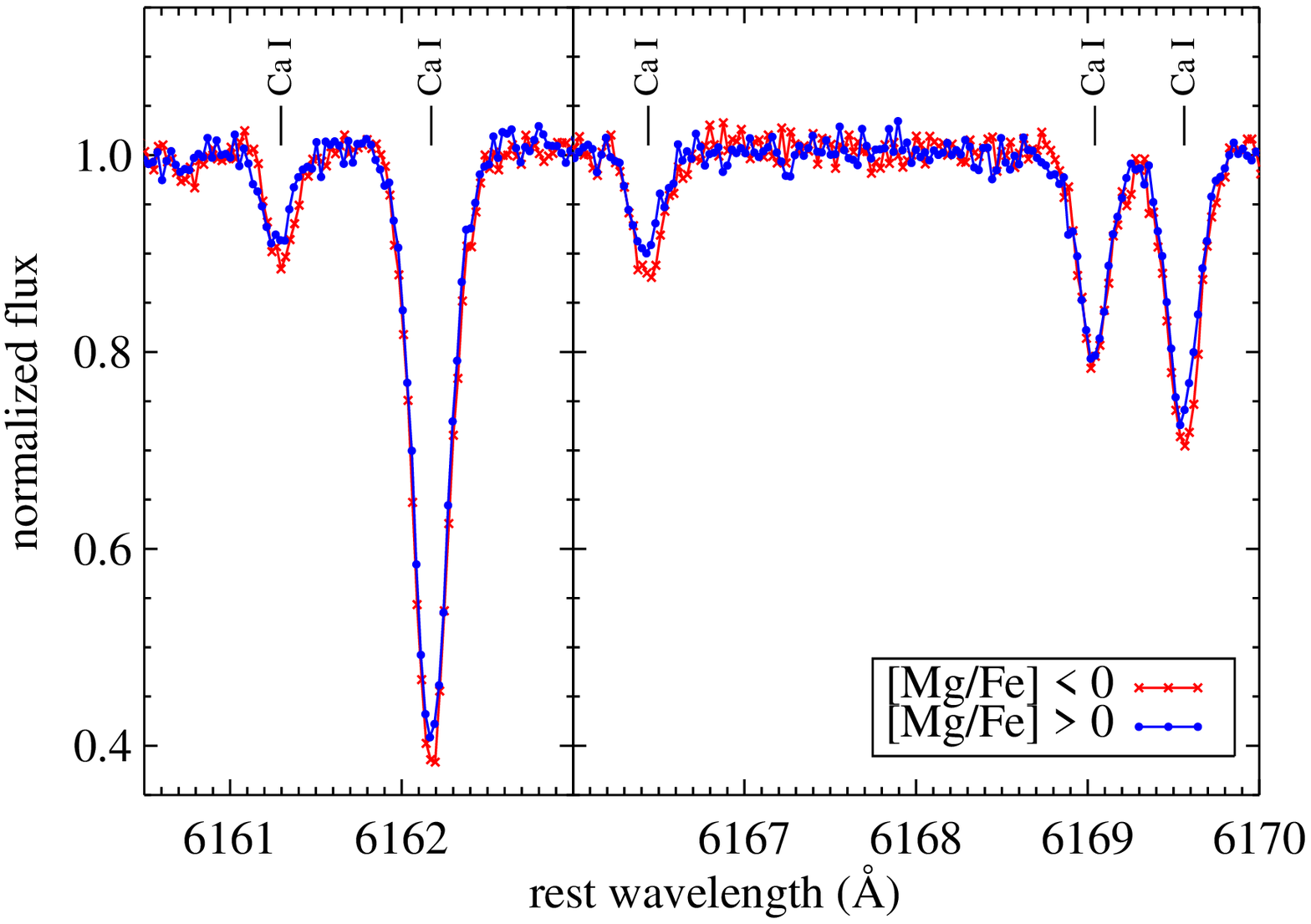}
\caption[]{Sums of sections of spectra for the Mg-normal (red crosses)
  and Mg-poor (blue points) NGC~2419 giants covering several
  \ion{Ca}{1} lines near 6165~\AA\@.  The \ion{Ca}{1} lines are almost
  identical, with the lines of the Mg-poor group being slightly
  deeper.  The sum for the Mg-normal stars omits S223, the most
  luminous and reddest NGC~2419 giant.
\label{figure_6165ca}}
\end{figure}

Figs.~\ref{figure_5197} to \ref{figure_6165ca} show sums of the
spectra of the Mg-poor RGB stars in NGC~2419 and those of the
Mg-normal stars to illustrate the contrast between the two groups.
The Mg-normal star S223, the brightest and reddest cluster member, has
been omitted from all the sums.  S223 has broader metal lines and
extremely strong H$\alpha$ emission.  Note that the mean \teff\ of the
Mg-poor stars is $\sim$100~K hotter than that of the Mg-normal stars
(see Fig.~\ref{figure_cmd}).  The specific features shown are (5) a
set of \ion{Fe}{1} lines, with one \ion{Fe}{2} line, to illustrate
that there is no evidence of a variation in Fe abundance, (6) the
region of the 5528~\AA\ \ion{Mg}{1} line which includes a \ion{Sc}{2}
line, (7) the Na doublet at 5685~\AA, to show that the Na
abundance is low in both groups, (8) an even weaker Na doublet near
6160~\AA, and (9) a set of \ion{Ca}{1} lines near 6165~\AA\@.  The
last of these figures shows that the \ion{Ca}{1} lines in the Mg-poor
group are only slightly stronger than those in the Mg-normal group of
NGC~2419 giants.  If the differences in line strength were due to
temperature alone, the lines would be weaker.  Hence, there is a
difference in the average Ca abundance between the two groups.

The extremely low [Mg/Fe] abundances we have determined for the Mg-poor group of 5
giants in NGC~2419 are very unusual.  Such low values of [Mg/Fe] can
be found only in the most metal-rich stars in dwarf spheroidal
galaxies \citep{letarte10,cohen_umi,kirby11}.  It is never found in
stars with the metallicity of NGC~2419 (${\rm [Fe/H]} = -2.1$).  For
example, the 122  stars of the 0Z project have a
median [Fe/H] of $-2.9$~dex, and 10\% of the sample
has [Fe/H] $> -2.3$~dex.  The
 lowest value of [Mg/Fe] in this sample of Galactic halo field EMP
 candidates is $-0.23$~dex, with normal [K/Fe].
Note that
only three of 122 stars in their sample have ${\rm [Mg/Fe]} < 0$
(Cohen et al., in preparation).  Furthermore, low values of [Mg/Fe] in
halo and dwarf galaxy stars are always accompanied by low abundances
of other $\alpha$ elements, such as Si and Ca, where they signify the
increasing role of (delayed) contributions from Type~Ia supernovae, which are
very effective at producing Fe-peak elements, to the chemical
inventory.  The situation in NGC~2419 is completely different,
especially because Fe and other Fe-peak elements show no variation in
NGC~2419.
   
      
\subsection{Consequences of the Large Range in Mg}   

Ignoring H and He, the most abundant elements in a scaled solar
mixture are C, N, O, Ne, Mg, Si, S, and Fe. Of these,  C, N, O, Ne, and S all
have high first ionization potentials, $\chi > 10$~eV.  Thus, in the
atmospheres of cool stars, Mg, Si, and Fe are the dominant sources of
free electrons, and of these three, Mg has the lowest first ionization
potential, and hence may be the most important.  We have established
that there is a substantial population of luminous giants in NGC~2419
with a very large deficiency of Mg.  Given the potential importance of
Mg to the structure of the stellar atmosphere, we must consider the
potential implications of such a large deficiency.  We investigated
the structure of the model atmospheres, in part, in response to the
suggestion by \citet{mucciarelli} that the induced change in electron
pressure, $P_e$, affects the CaT enough to produce the observed
dispersion in line strengths in NGC~2419.

If the Mg abundance were to be increased from some initial level, one
would expect $P_e$ to rise, and to continue rising as the Mg abundance
is increased further.  However, decreasing the Mg abundance from some
initial level does not produce the same behavior, as once Mg is
sufficiently depleted, it will no longer be an effective electron
donor compared to other sources of electrons, and any additional
decrease in $P_e$ will occur much more slowly.  Thus, whatever the
effect may be of the strong Mg depletion seen in the Mg-poor group of
RGB stars in NGC~2419, we should not expect a very large range in
behavior arising from the large range in the depletion of Mg within
the Mg-poor population in NGC~2419.

One potential effect of a major change in the Mg abundance is that the
position of the RGB in the CMD may shift depending on the Mg abundance
to some extent.  \cite{vandenberg12} recently evaluated the effects of
altering the abundance of a single element drawn from a large list of
suitable elements, including Mg.  They found that increasing Mg by
0.4~dex at ${\rm [M/H]} = -1.0$ substantially increases the opacity in
the atmosphere and substantially shifts the RGB locus redder.
However, at ${\rm [M/H]} = -2$, the shift becomes much smaller, as can
be seen by comparing their Fig.~7 to their Fig.~10 (the resulting
opacity change) and their Fig.~15 to their Fig.~16 (the RGB positions
in a CMD for these two values of [M/H] and for increases in Mg, Si, or
Ca).  In NGC~2419 where ${\rm [M/H]} \sim -2$, the [Mg/Fe] value is
deficient well below the normal $\alpha$-enhanced value to one with
[Mg/Fe] between $-0.9$ and $-0.2$~dex.  It is clear from the
calculations of \cite{vandenberg12} that the position of the RGB will
{\it{not}} be perceptibly altered in this situation.

We initially expected, based on the substantial shifts in the RGB
found by \cite{vandenberg12} at higher metallicity, that the Mg-rich
and Mg-poor populations in NGC~2419 would be separated in the RGB with
the Mg-poor population lying somewhat to the blue of the main RGB\@.
However, CMDs using both $V-I$ and $V-J$ (see Fig.~\ref{figure_cmd})
show this is not the case.  Understanding that no such shift is
expected to happen for the specific case of NGC~2419 relieves our
initial concern.  It would maintain the validity of our normal methods
of stellar parameter determinations, especially \teff, which rely on
broad band colors.  (However, we do not use colors here, instead
relying on $V$ alone.)  It also gives specific guidance for the special
case of the star S1673, which is the most Mg-depleted star in our
sample.  It lies somewhat to the blue in the $V$, $V-I$ CMD, but less
so in $V$, $V-J$.  Its location blueward of the RGB is either an
unexpectedly large error in \citeauthor{stetson05}'s
(\citeyear{stetson05}) visual photometry or is a reinforcement of our
earlier suggestion that S1673 is an AGB star.

A second issue to consider is the effect of any decrease in $P_e$ on
the formation of spectral lines.  Because these are luminous cool
giants, the temperatures are low, and most elements are mostly
neutral.  As a result, any decrease in $P_e$ from a depletion of Mg
will have little effect on the neutral lines, but the number density
of the singly ionized species will rise.  Is this the reason that the
near-IR triplet, which is a line of \ion{Ca}{2}, is enhanced in the
Mg-poor population?  The key question is whether $P_e$ is affected by
the decrease in Mg abundance within the stellar atmosphere, or whether
the decrease in $P_e$ at the metallicity of NGC~2419 is so small that
there is no obvious change.

To this end, we calculated some model atmospheres at the stellar
parameters characteristic of our NGC~2419 sample with Mg enhanced and
depleted by 0.7~dex in each case.  These tailored model atmospheres
use as a base the $\alpha$-enhanced models from \cite{castelli04},
which have scaled solar abundances but with ${\rm [\alpha/Fe]} =
+0.4$, where all elements with even atomic number from O through Ti
are considered $\alpha$-elements.  From this base composition the Mg
abundance is perturbed up or down by a factor of 5 (0.7~dex) to
construct new model stellar atmospheres.

\begin{figure}
\centering
\includegraphics[width=\linewidth]{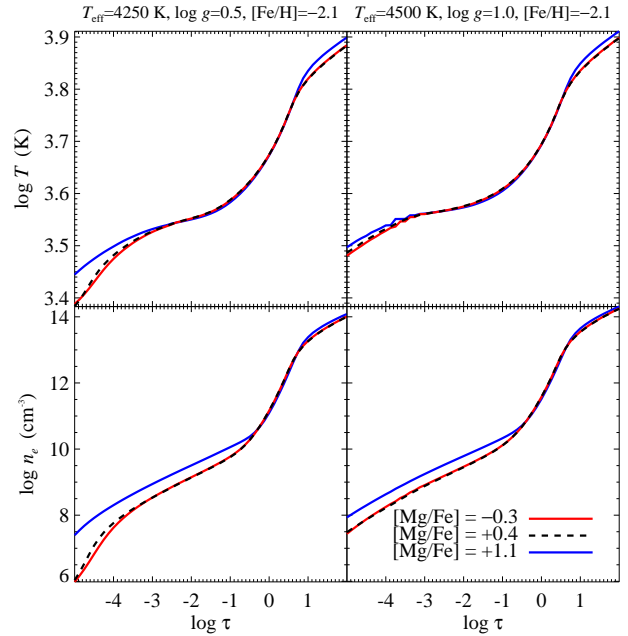}
\caption[]{$T(\tau)$ and $n_e(\tau)$ are shown for stellar model
  atmospheres from \citet{castelli04} that represent the Mg-normal
  population in NGC~2419 together with tailored models for Mg enhanced
  or depleted by a factor of 5.  Note that $n_e$ is identical for the
  Mg-normal and Mg-depleted models deeper than $\tau \sim 10^{-4}$,
  while the very highly Mg-enhanced models with ${\rm [Mg/Fe]} \sim
  +1.1$ do show the expected trend of an increase in $n_e$ at all
  optical depths deeper than $\tau \sim 10^{-4}$.
\label{figure_pe_tau}}
\end{figure}

Fig.~\ref{figure_pe_tau} shows $n_e$ and $T$ as a function of $\tau$
for the base model atmospheres from \cite{castelli04} (with ${\rm
  [Mg/Fe]} = +0.4$), and for those with Mg further enhanced or
depleted by a factor of 5 for two sets of stellar parameters, both
with ${\rm [M/H]} = -2$.  The highly Mg-enhanced model (with ${\rm
  [Mg/Fe]} = +1.1$) shows the expected behavior: a strong enhancement
of $n_e$ at all depths with $\tau > 10^{-4}$ compared to the base
model.  However, the Mg-depleted model closely follows the base model,
implying that with such a large depletion of Mg, and given the low
overall abundance of this GC of $-2$~dex, Mg is no longer an important
source of free electrons.  We might have expected this from the
detailed isochrone calculations of \cite{vandenberg12}, but it is
gratifying that this is verified by our stellar model atmosphere
calculations.

This agreement for $\tau > 10^{-4}$ suggests that our detailed
abundance analyses which have been carried out using the base
\cite{castelli04} models will be valid for both the Mg-normal and the
Mg-poor populations in NGC~2419, with the possible exception of lines
formed higher in the atmosphere than $\tau = 10^{-4}$.  None of the
lines included in the detailed abundance analysis are strong enough
for this to be the case.  The only relevant spectral features that may
be strong enough to be formed at so near the surface are the 
cores of the near-IR
triplet of \ion{Ca}{2} which were used in our DEIMOS analysis
(\citeauthor*{deimos}) to suggest the possibility of a variation in
the Ca abundance within NGC~2419.  We note that any change in $P_e$
does not affect the pressure broadening for lines strong enough to
show damping wings as it is dominated by interactions with neutral H
atoms (van der Waals broadening).  It is the possible effect on the
ionization balance for Ca via change in $P_e$ that is of concern.

\begin{figure}
\centering
\includegraphics[width=\linewidth]{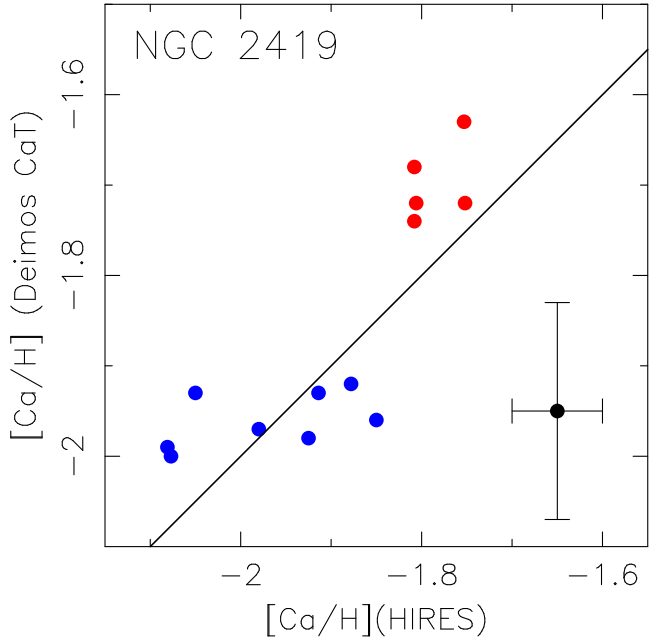}
\caption[]{A comparison of the Ca abundance derived here from lines of
\ion{Ca}{1} for our sample of 13 RGB stars in NGC~2419 with
Keck/HIRES spectra compared to that deduced by C10 from the infrared
Ca triplet of \ion{Ca}{2}. A line indicating equality is shown.  A typical
error bar is given in the lower right corner. Red and blue points
  denote Mg-poor and Mg-rich giants respectively.
\label{figure_hires_cat}}
\end{figure}

However, we reject the suggestion of \citet{mucciarelli} that this
is an important issue for NGC~2419 based on the good agreement between
the abundance of neutral Ca lines presented here vs that from
the near-IR triplet of \ion{Ca}{2} presented by C10
(see Fig.~\ref{figure_hires_cat}), as well as
from the evidence regarding the behavior of tailored model atmospheres
as the Mg abundance is varied shown in Fig.~\ref{figure_pe_tau}.
The prediction that there is no shift in the ionization equilibrium for
the case of interest here means that achieving ionization equilibrium
with the set of lines from the Keck/HIRES spectra used here is an
important constraint that can be used to validate our detailed
atmosphere abundance analyses.
If NGC~2419 were more metal-rich by a factor of 4 or more, 
or Mg in the bulk of its population
was more enhanced than we have established it to be, the consequences
of differing Mg abundance between the two populations in this
GC on $P_e$ would become an important issue.

Although the Mg-poor stars in NGC~2419 also show a strong enhancement
of potassium, K is less abundant than Mg in the solar mixture by a
factor of $\sim$250, and so its enhancement by a factor of less than
10 in these stars will not produce any significant effect on the
structure of the stellar atmosphere.

\section{Is there a Spread in Abundances within NGC~2419?}
\label{section_abund_spread}

We have established that there are two groups of RGB stars in
NGC~2419: those that appear like normal GC RGB stars with ${\rm
  [Mg/Fe]} \sim +0.3$ and a second group with extremely low Mg
abundances (the Mg-poor group of 5 stars, with [Mg/Fe] ranging from
$-0.2$ to $-0.9$~dex).  Furthermore, the Mg-poor stars are those that
have high Ca(CaT) from our Keck/DEIMOS study (\citeauthor*{deimos}).
There is an anti-correlation with the K abundance such that the
Mg-normal giants have ${\rm [K/Fe]} \sim +0.4$, a value consistent
with that of other GCs and metal-poor halo stars (see, e.g.,
\citeauthor{first_starsv} \citeyear{first_starsv} or
\citeauthor{emp_dwarfs} \citeyear{emp_dwarfs}, in which there are no
non-LTE corrections), while the Mg-poor giants have ${\rm [K/Fe]} \sim
+1.1$, a factor of 5 higher than the Mg-normal stars.  The last
element with detectable variations is Sc.  As shown in
Table~\ref{table_mean_abund} and in Fig.~\ref{figure_5197}, there is
no credible evidence for variation of Ti or any heavier Fe-peak
element, including Fe itself.

\subsection{Mg and K}

Are the very large abundance variations seen among the NGC~2419 luminous RGB
stars for Mg and K real?  First, we discuss the case of Mg.  There are
three to five detected \ion{Mg}{1} lines per star.  The range in Mg
abundance among the giants in our sample in NGC~2419 exceeds a factor
of 10.  There is no way that this can arise from a problem in the
abundance analysis procedure.  The non-LTE corrections for \ion{Mg}{1} lines
are small.  \citet{mg_nonlte} found that the typical correction for
metal-poor giants is about $+0.2$~dex and not particularly sensitive
to atmospheric parameters.  Previous non-LTE computations reached
similar results for disk stars \citep{mishenina04} and for a range of
stellar types and metallicities, down to ${\rm [M/H]} = -2$
\citep{shimanskaya00}.  All three of these studies used many of the
same \ion{Mg}{1} transitions that we used in our abundance analysis.
We conclude that the variations in Mg abundance between the Mg-normal
and Mg-poor stars must be real.

The situation with K is less clear.  The variation of $\epsilon({\rm
  K})$ in NGC~2419 (spanning a range of 1.4~dex) is large enough so
that any inaccuracies caused by the stellar parameters or the analysis
code are too small to produce the observed spread.  However, the only
lines of K that can be observed at optical wavelengths are the
resonance doublet at 7665 and 7699~\AA\@.  Because the former is
embedded deeply within a very strong terrestrial absorption band of
O$_2$, it is practical to measure only the 7699~\AA\ line.  Given that
this is a fairly strong resonance line, non-LTE corrections need to be
considered.

Non-LTE corrections for the 7699~\AA\ line of \ion{K}{1} have been
calculated by several groups
\citep[e.g.,][]{ivanova_k,takeda_knonlte,mg_nonlte}.  The non-LTE
corrections are negative, and range from $-0.1$ to $-0.9$~dex (see
Fig.~6 of \citeauthor{ivanova_k} \citeyear{ivanova_k}).  They vary
strongly with \teff\ and with metallicity.  \citet{ivanova_k} wrote
that ``the non-LTE corrections can vary strongly as functions of the
model atmosphere parameters, which can sometimes be a source of
substantial errors, even when comparing potassium abundances for stars
of very similar type.''  It may be possible, with some contortions, to
reproduce the behavior of K between the Mg-normal and Mg-poor stars in
NGC~2419 with non-LTE corrections alone, but it does not seem likely.

Although the surveys of GC and halo field stars carried out prior to
2004, including the extensive work of the Lick-Texas group
\citep[e.g.,][]{kraft94} and of Cohen and her collaborators
\citep[e.g.,][]{cohen_m13}, did not include the
\ion{K}{1} lines due to limitations on spectral coverage, more recent
work has found a small number of other metal-poor Galactic giants that
show the very high K abundances of the Mg-poor giants in NGC~2419.
\cite{takeda_highk} found two such stars in a survey of 15 RGB stars
in three GCs.  These two stars stick out in the same way as the K of
the Mg-poor stars stick out in NGC~2419.  One of these stars, M13
III--73, which has a [K/Fe] abundance 0.6~dex higher than the rest of
the M13 sample, has been analyzed in detail by \cite{kraft92} and
\cite{pilachowski96}\footnote{We are trying to obtain a better HIRES
  spectrum of M13 III--73; it should be in hand shortly.}.  They find
that ${\rm [Mg/Fe]} = +0.25$, a normal value for an $\alpha$-enhanced
GC star.
\citet{first_starsv} found CS~30325--094, an EMP giant with ${\rm
  [Fe/H]} = -3.3$, to have ${\rm [K/Fe]} = +0.72$, with normal
$\alpha$-enhancement, and [Mg/Fe] and [Ca/Fe] both at +0.38~dex. (Note
for future reference that [Sc/Fe] in this star is rather high at
+0.33~dex.) 

In summary, there are a few stars with [K/Fe] similar to those of the
Mg-poor giants in NGC~2419, but they are not Mg-poor, in general.
This suggests that the process generating K is not always tied to that
producing the Mg-poor anomaly.

\begin{figure}
\centering
\includegraphics[width=\linewidth]{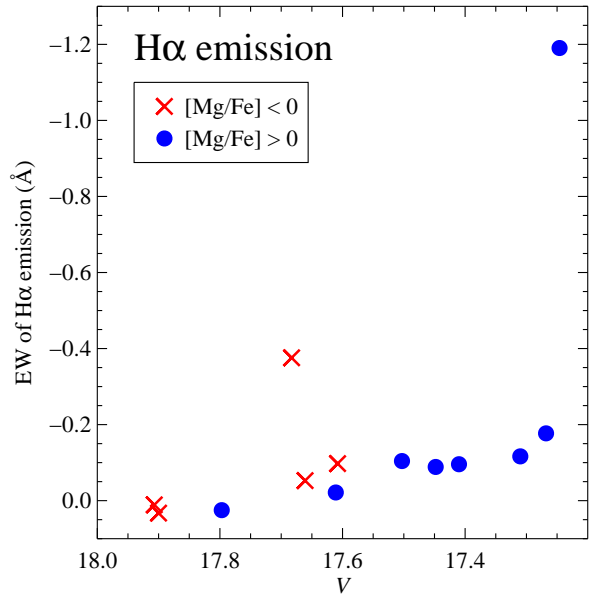}
\caption[]{The sum of $W_{\lambda}$ of emission in the blue and red
  wings of H$\alpha$ as a function of $V$ for our sample of luminous
  giants in NGC~2419.  The red crosses indicate the Mg-poor stars, and
  the blue points represent the Mg-normal stars.
\label{figure_halpha}}
\end{figure}

\cite{takeda_highk} suggested that some anomalously strong K resonance
lines are caused by exceptional cases of strong peculiar velocity
fields in the upper layers of the atmosphere and do not reflect the
true K abundance of the star.  To explore this possibility in our
NGC~2419 sample, we measured the radial velocity of the \ion{K}{1}
7699~\AA\ line and compared it to that measured from other lines.  We
find that these agree for all 13 stars in our sample to within
0.5~\kms\ for all the stars except one, where the difference is
0.8~\kms\@.  Furthermore, we used the H$\alpha$ emission as a measure
of chromospheric activity and possible mass loss.  \citet{cohen_76}
discovered that weak H$\alpha$ emission is common in the brightest red
giants in GCs.  She interpreted the emission as mass loss from the
bloated atmospheres of the giants.  However, the implied mass loss
rate was large.  \citet{dupree84} later found from models of the
atmospheres of red giants that static chromospheres can explain both
the emission and the blueshift of H$\alpha$ in bright, red giants.
$W_{\lambda}$ of the sum of the blue and red emission wings in
H$\alpha$ for luminous RGB stars in NGC~2419 are shown as a
function of $V$ (our proxy for \teff) in Fig.~\ref{figure_halpha}.
Some of the sample stars show weak emission, but two stand out.  The
strongest H$\alpha$ emission, which is extremely strong, is shown by
S223, which is the most luminous and reddest RGB star in NGC~2419.  It
has obvious strong emission in both the red and blue wings of
H$\alpha$ and H$\beta$ and also in the blue wings of H$\gamma$ and
H$\delta$, as well as having broader metal lines than the other sample
giants.  The other case of strong emission is S1673, the star
suspected to be on the AGB\@.  Its emission is much weaker than S223
(see Fig.~\ref{figure_halpha}) but stronger than the other NGC~2419
giants in our sample.  The Mg-poor stars behave no differently from
the Mg-normal RGB stars in NGC~2419 in terms of their H$\alpha$
emission. The mean radial velocity of H$\alpha$ in the 5 Mg-poor stars
differs from that of the metallic lines by only $-0.3$~\kms;
the same value for   the 8 Mg-normal stars in NGC~2419 is $-0.6$~\kms.  
The lithium line at 6707~\AA\
cannot be detected in the summed spectra of either group of NGC~2419 giants. 
We therefore find untenable the suggestion that the high K
seen in the Mg-poor giants in NGC~2419 arises from velocity fields in the outer
layers of the stars.  Furthermore, at the metallicity of NGC~2419,
even for the Mg-poor, K-strong stars, the 7699~\AA\ line of K is not
very strong, and most of it is not formed extremely high in the
stellar atmosphere.

\subsection{Si, Ca, and Sc}

We now turn to elements which appear to show smaller variations within
the stellar population of NGC~2419.  According to
Table~\ref{table_mean_abund} these are Si, Ca, and Sc.  (The neutron
capture elements may show a weakly significant dispersion, and we
discuss them below.)  Since there is the most information about Ca,
we discuss it first.

The existence of Ca variations was first suggested by
\citeauthor*{deimos} based on their analysis of the strengths of the
8542 and 8662~\AA\ lines (the two stronger lines of the CaT) from
moderate resolution Keck/DEIMOS spectra of a large sample of stars.
The key issue is whether this spread is caused by a real star-to-star
abundance variation, or whether, since these features arise from
singly ionized Ca, it is a consequence of a decrease in $P_e$ due to
the very low Mg abundances in the Mg-poor cluster giants.  The present
analysis (see Table~\ref{table_mean_abund}) which is based on much
higher resolution spectra from which typically 15 \ion{Ca}{1} lines
can be measured, also suggests a spread in Ca.  Furthermore,
\citeauthor*{deimos} also presented the results of spectral syntheses
using the method of \citet{kirby10} on the DEIMOS spectra, which
specifically included \ion{Ca}{1} lines in the appropriate wavelength
region but excluded the CaT lines due to uncertainties in their line
formation.  Fig.~\ref{figure_hires_cat} compares the Ca abundance
derived here from \ion{Ca}{1} features in our Keck/HIRES spectra with those
derived by \citeauthor*{deimos} from the infrared triplet lines of
\ion{Ca}{2}. The agreement is quite satisfactory; both show a small spread
of $\sim$0.2~dex in Ca abundance with the Mg-poor population having
a higher Ca abundance than the Mg-normal population in NGC~2419.


Variations in Ca abundance within a GC are quite unusual.
\cite{carretta_ca} placed very tight limits on any variation in [Ca/H]
of not more than 0.03~dex in a sample of 17 GCs.  The only previously
known GCs that show such variations are those suspected of being
remnants of formerly accreted dwarf galaxies, such as $\omega$~Cen.
From both spectroscopy and photometry, $\omega$~Cen has been known for
more than 30 years to have a wide intrinsic range in [Ca/H], [Fe/H],
and many other elements, extending over a range of $\sim$1.3~dex with
multiple peaks in the metallicity distribution \citep{norris96}.
Other GCs with spreads in [Fe/H] include M22 \citep{marino11}
and NGC~1851 \citep{carretta_1851}.

One of the unresolved puzzles of \citeauthor*{deimos} was the contrast
between the constancy of [Fe/H] within the DEIMOS sample and the
spread seen in Ca based on the strength of the near-IR triplet of
\ion{Ca}{2}.  The present HIRES analysis confirms this surprising result,
namely that there is no detectable spread in [Fe/H], yet there is a
small one in [Ca/H] in NGC~2419.  Those ``globular clusters'' such as
$\omega$~Cen which do show variations in [Ca/H] within their stellar
populations also show comparably large variations in [Fe/H].  The
absence of a spread in Fe makes NGC~2419 unique in the details of its
chemical inventory.

We considered the range of variation in Ca(CaT) as compared to the
HIRES \ion{Ca}{1} result.  The means for [Ca/H] of the Mg-poor and the
Mg-normal stars inferred from their HIRES spectra differ by 0.18~dex
(see Table~\ref{table_mean_abund}) while the difference for the same
13 stars for Ca(CaT) from our DEIMOS spectra (see Fig.~\ref{figure_hires_cat})
is somewhat larger, 0.27~dex.  Nonetheless, the two independent values
for the difference in Ca abundance between the Mg-poor and Mg-normal
stars in NGC~2419
agree within the uncertainties.  Furthermore, the Mg-poor and
Mg-normal groups both show internal dispersions in Ca abundance that
are considerably smaller than the difference between them.  This
applies to both the HIRES sample of 13 stars and the larger DEIMOS
sample of \citeauthor*{deimos}.

Non-LTE corrections for Ca lines have been calculated by several
groups, most recently by \cite{spite2012}.  At the metallicity of
NGC~2419 they are essentially zero and hence negligible for the
subordinate \ion{Ca}{1} lines, but that is not the case for the
4226~\AA\ resonance line.  Since the S/N at 4226~\AA\ in our spectra
is poor, we do not use the resonance line anyway.  As a result,
non-LTE effects are not an issue for the set of \ion{Ca}{1} lines that we
used for our present high-resolution study of luminous RGB stars in
NGC~2419.

We therefore conclude that there is a real, but small, variation in Ca
abundance between the Mg-poor and Mg-normal luminous giants in
NGC~2419.  The Mg-poor stars have a higher [Ca/H] abundance by
$\sim$0.2~dex.

[Si/Fe] also shows a low amplitude anti-correlation with [Mg/Fe] such
that the Mg-poor stars have values $\sim$0.2~dex higher than the
Mg-normal stars in NGC~2419.  This can be seen in
Fig.~\ref{figure_5680na}.

The mean [Sc/Fe] we derive from our HIRES spectra of stars in NGC~2419
is 0.45~dex higher in the Mg-poor stars than in the Mg-poor stars.
This difference is easily seen in the composite summed spectra of the
Mg-poor and Mg-normal giants shown in Fig.~\ref{figure_5528} and in
Fig~\ref{figure_5680na} as well as in the plot of [Sc/Fe] vs.\ [Mg/H]
shown in Fig.~\ref{figure_abund_mg}.  We assert that the Sc abundance
is noticeably different in the mean between the Mg-poor and Mg-normal
groups of giants.

We note again that the Fe abundance is constant across both Mg-poor
and Mg-normal giants in NGC~2419 to within 0.1~dex, as is shown in
Table~\ref{table_mean_abund} and in Fig.~\ref{figure_5197}.
\cite{carretta_fe} established strong upper limits on any star-to-star
variation in [Fe/H] in a large sample of Galactic GCs.  Only those GCs
that are widely believed to be the remnants of accreted dwarf galaxies
show star-to-star variations in [Fe/H]\@.

\subsection{The Neutron Capture Elements \label{section_ncapture} }

Five of the heavy neutron capture elements 
(Y, Ba, Ce, Nd, and Eu)
are detected in 10 or
more of the 13 RGB stars in our HIRES sample for the GC NGC~2419,
all as singly ionized species.
The dispersion of [X/FeII] for these 6 elements is reasonably
small considering that with the exception of Ba, for each of these
species we have detected 
only a few weak lines redder than 4200~\AA.  Ba has four
strong lines in the spectral region studied, most of which were
detected in all the sample stars.  Eu has a strong line
at 4129~\AA, but our spectra have low S/N there.  So 
the abundance of Ba is the most reliable among these elements.
The ratio of [Eu/Ba] for our sample of 13 RGB stars in 
NGC~2419 is +0.33~$\pm$0.11~dex, comparable to that seen in other
metal-poor GCs \citep[see, e.g.][]{gratton04}.

\begin{figure}
\centering
\includegraphics[width=\linewidth]{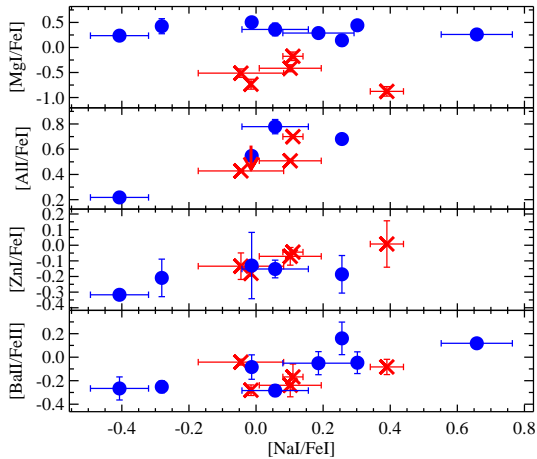}
\caption[]{Abundance ratios for selected species with respect to Fe,
  using \ion{Fe}{1} or \ion{Fe}{2} as appropriate, are shown as a
  function of [Na/Fe] for our sample of 13 giants in NGC~2419 with
  HIRES spectra.  The ratios selected are sensitive to proton burning
  chains operating among Na, Mg, and Al, and to a possible $s$-process
  contribution from intermediate-mass AGB stars.  Red and blue points
  denote Mg-poor and Mg-rich giants respectively.
\label{figure_abund_na}}
\end{figure}

Table~\ref{table_mean_abund} demonstrates that
there is no apparent difference exceeding 0.1~dex between abundances of any
of these 6 elements between the Mg-poor and the 
Mg-normal population in NGC~2419.  However, as is shown
in Fig.~\ref{figure_abund_na}, there is a weak correlation
between [NaI/FeI] and [BaII/FeII], and an even stronger 
correlation of  [NaI/FeI] and [ZnI/FeI].  There is also a hint
of a correlation between Al and Na, but the data are too
sparse to be certain since the S/N at the strong Al~I line at 3961~\AA\
is too low to permit the use of that feature, and we must rely
on the much weaker doublet at 6690~\AA.
No such correlation with
any other element heavier than Fe with sufficient data  
was seen in our NGC~2419 sample. We may consider the
Na abundance as a proxy for the typical mode of multiple populations 
involving proton-capture at high temperatures among the light elements
seen in essentially all GCs as an anti-correlation between Na and O
abundances.  If we view the high Na abundance stars as those of
the typical second generation, then we might consider the high Zn and Ba
abundances as indicating a contribution from the $s$-process,
such as was first seen in NGC~1851 by \cite{yong08}, who found
that the Zr and La abundances of a small sample of stars were
correlated with Al, and anti-correlated with O.
\cite{carretta_1851_ba} present
more recent results with a larger sample and find correlations
between the Al and Ba abundances in this GC.

Thus NGC~2419 is unique among the GCs in that it has two distinct
manifestations
of multiple populations.  The first is the classic proton burning at high $T$
as manifested  by correlations and anti-correlations among the light
elements, seen in NGC~2419 as a range of Na abundances, and the related $s$-process
contributions of Zn and Ba.  The second is
the strong Mg - K anti-correlation, which appears
to have a completely separate origin.

\section{Potential Causes of the Anomalies Seen in NGC~2419} 

In the previous section, we reviewed the evidence for anomalies in the
chemical inventory of NGC~2419 and demonstrated that they are almost
certainly real.  They are not artifacts of analysis problems or
non-LTE issues (except possibly for the enhancement of K, but probably
not), and they require a nucleosynthetic explanation.  We discuss
these in order of the magnitude of the anomaly.

As reviewed by \cite{gratton04}, it is now well established that
all GCs contain (at least) two generations of stars: the primordial
generation, plus a second one whose light elements (C, N, O, Na, Mg,
Al) show evidence for proton burning beginning with C and O burning
into N, Ne burning into Na, and Mg burning into Al.  Correlations and
anti-correlations found among the light elements in GCs have
demonstrated this high-temperature proton-burning occurs in the
progenitors of the GC stars. A major study of these issues is
summarized by \cite{carretta_nao}.

Furthermore, the light element abundance variations persist all the
way down the RGB to the SGB and to and even below the main sequence
turn-off \citep{briley96}.  Thus, they cannot be attributed to stellar
evolution within a single star, but must involve material processed in
more massive stars, then ejected, with the usual suspects being
intermediate mass AGB stars, as advocated by \cite{dantona02}
\citep[see, e.g.][for a more recent view]{dantona12}, or rapidly
rotating massive stars \citep{maeder06}.  A low-amplitude correlation
of Si with Al ($\sim$0.2~dex increase in [Si/Fe] for [Al/Fe]
increasing by 1.5~dex) is seen in a few GCs as well, e.g.,
\citet{yong_6752} and \citet{carretta_6752}, among others, implying
proton burning occurring in even hotter environments.

AGB stars are also strong sites for the $s$-process, and thus the
search for correlations with $s$-process element variations in GCs is
also important. There are some hints that extensive proton burning
producing very strong enhancements of Na (from Ne) and Al (from Mg)
also produces small amounts of $s$-process material leading to
correlations between small, marginally statistically significant
enhancements of Y, Zr, and Ba with Na enhancements, as well as with much larger Al
enhancements \citep{yong_6752}.

One might try to invoke a similar process to this to explain at least
part of the anomalies in NGC~2419.  However, Mg is a very abundant
element, and in normal GCs, burning 40\% of the original Mg will
produce an enhancement of a factor of 10 or more in the Al abundance
in the second generation stars.  The burning of 90\% of the Mg will
produce an enhancement of Al which is much larger than that seen in
the Mg-poor stars in NGC~2419.  However, Fig.~\ref{figure_abund_na}
demonstrates that while some of the usual correlations and
anti-correlations among the light elements present in NGC~2419, the
amplitude of the Al dispersion is by no means exceptional, as defined
by the behavior of a sample of 15 GCs studied by \cite{carretta_nao}.
It is especially puzzling that Mg itself does not correlate with Na,
Al, Ba, or any other element typically indicative of proton burning or
the $s$-process.

Furthermore, we have demonstrated that in the Mg-poor stars, K, Ca,
and Sc, elements well beyond Si, are also enhanced.  We rule out
proton burning among the light elements as an explanation for the
anomalies in the chemical inventory of NGC~2419 as it is impossible to
reach the required temperature outside supernovae.  Such burning may
well be going on at a very low level, but it is at best a minor
contributor to the bizarre behavior we are trying to explain.

To explain the strong depletion of Mg seen in the Mg-poor population
in NGC~2419 requires nuclear burning at high temperatures and beyond
the range of nuclear processing believed to occur at the bottom of the
surface convection zone in AGB stars.  Mg is produced during the CNO
cycle operating in the cores of massive stars, equivalent to
$2^{12}{\rm C} \rightarrow ^{24}{\rm Mg}$.  It is also produced
copiously in Type~II supernovae.  We have at present no explanation for the Mg-poor
population.

Potassium is the one element discussed in this section where there is
at least a semi-viable, non-nuclear explanation, namely non-LTE effects
(see \S\ref{section_abund_spread}).  But as discussed above, this is
rather contrived, and probably cannot be made to work.  Potassium is
much less abundant in scaled solar mixtures than even Al, so if one
tries to invoke proton burning cycles to produce the excess K seen in
the Mg-poor population, a much larger enhancement of K is predicted
than is observed.

K is primarily produced by oxygen burning in Type~II SNe, but, as
discussed by \cite{clayton}, its production depends heavily on the
progenitor mass and on the assumptions regarding fallback and when
material is ejected during the SN explosion.  Scandium is even rarer
than K in a scaled solar mixture, and its abundance in Type~II SN
ejecta depends crucially on how far oxygen burning has proceeded in
material before it is ejected.  Varying only the progenitor mass, the
Type~II SN yields of \cite{nomoto06} show a peak in production of both
K and Sc with respect to Ca for a Type~II SN progenitor mass of
between 18 and 20~$M_{\odot}$ depending on the initial metallicity
(zero or low) of the SN progenitor.  Given this, it may be possible,
by tinkering with the characteristics of Type~II SN explosions, to
produce highly varying fractions of K and of Sc in the ejecta. Since
the yields of K and Sc given by \cite{nomoto06} vary more or less
together, one might expect to see correlated abundances of K and Sc, as
is the case in NGC~2419.

The dominant isotope of Si is $^{28}$Si, which can be assembled from 7
nuclei of $^4$He.  It is very tightly bound and is the primary product
of O burning in the cores of massive stars.  The dominant isotope of
Ca is $^{40}$Ca, which has the magic number 20 of both protons and
neutrons. Thus, it is very stable compared to its neighbors in the
periodic table of the elements.  It can be assembled from 10 nuclei of
$^4$He.  It, too, is produced primarily in Type~II SNe during O
burning.  The abundances of these very stable elements in Type~II SN
ejecta are less sensitive to the details of the explosion than those
of K or Sc. This may be why the mean differences in abundance between
the Mg-poor and Mg-normal populations in NGC~2419 for Si and Ca are
just 0.2~dex between the two groups.

Type~II SN nucleosynthesis models including ejection mechanisms,
fallback, and mixing within the ejecta can successfully explain the
chemical inventory of (most) EMP Galactic halo stars
\citep{kobayashi06,tominaga07,heger10}. If
one wishes to invoke peculiar Type~II SN explosions to explain the
anomalies in the chemical inventory of NGC~2419, since it is (even
now) a very massive GC, just one peculiar Type~II SN may not eject
enough material to produce a population of Mg-poor stars which
comprises $\sim$30\% of the present cluster stars.  Speculation that 
multiple, peculiar Type~II SNe occurred in NGC~2419 and in no other
known GC seems rather ad hoc and therefore unsatisfactory.

\section{Broader Implications Of the Large Mg Variations in NGC~2419} 

The depletion of Mg among the Mg-poor stars we have found in NGC~2419
is unprecedented among metal-poor stellar systems of any age or total
mass.  \cite{vandenberg12} calculated the effect of single-element
enhancements on GC isochrones.  At the low metallicity (${\rm [Fe/H]}
\sim -2.1$) of this GC, they found very small changes on the RGB
position in the CMD\@.  At higher metallicity, \cite{vandenberg12} found
substantial changes in the positions of the RGB in a simple stellar
population such as GC\@.  Furthermore, augmentations in Mg (or Ca)
relative to Fe would have even larger consequences than depletions.
The same holds true for the effects induced by a change in the
abundance of a specific element with regard to the line strengths of
individual spectral features, both those of the element involved, and
those of other elements through the effect of a change in $P_e$ and
hence a change in ionization ratios.  Spectral features originating
from an ionization stage which contains only a small fraction of the
total atoms of the relevant element can be significantly altered in
strength.  The RGB plus AGB dominate the total light at optical and IR
wavelengths in old stellar populations.  In these cool stars, it is
the population and potentially the line strength of the neutral
vs.\ the singly ionized species that may be affected.

If large divergences in the abundance of a single (abundant, low first
ionization potential) element from the scaled solar ratio or the
normal $\alpha$-enhanced ratio do occur in simple stellar systems,
then the consequences for the study of more distant stellar systems,
where only the integrated light can be observed, may be profound.  For
example, the calibrations relating [M/H] and the Mg triplet line
indices for the Lick indices \citep{worthey97,puzia05}, widely used to
interpret moderate resolution spectra of galaxies and GCs beyond the
Local Group, will be altered.  Substantial star-to-star variation
within a GC of the abundances of crucial elements could mimic a
variation of overall metallicity or an age spread.  By increasing the
number of parameters that must be considered, possible variations in
the abundances of key individual elements add considerable complexity
to the interpretation of the their CMDs.

So far only NGC~2419 shows such behavior in its chemical
inventory, and since this is such a metal-poor GC, the effects on its
CMD are very small, and the effects on its spectrum only appear in the
lines of elements that are actually abnormal in their abundances.
While finding many more such cases, especially at higher metallicity,
would be very interesting, for the sake of our entire knowledge base
of the composite light of simple stellar systems, we must hope that
such cases are very rare.

\section{Summary}  

Our initial work on the extremely distant and massive outer halo GC
NGC~2419 (\citeauthor*{deimos}) used moderate resolution spectra from
Keck/DEIMOS\@.  We suggested the presence of a star-to-star spread in
[Ca/H] but no detectable spread in [Fe/H] based on an analysis of the
strong \ion{Ca}{2} near-IR triplet and spectral synthesis of weaker,
neutral lines.  We then proceeded (\citeauthor*{cohen11}) to obtain
high-dispersion spectra of a sample of stars, most of which appeared
to be normal, $\alpha$-enhanced red giants similar to those found in
most GCs.  But \citeauthor*{cohen11}'s sample also contained one very
peculiar star, S1131, which showed very depleted Mg and highly
enhanced K and was apparently Ca-rich as well.  To follow this up, in
this paper we presented abundance analyses of 6 new RGB stars
in NGC~2419, most selected to be among the most apparently Ca enhanced
in the study of \citeauthor*{deimos}.

We found that there are two groups of stars in NGC~2419, one of which
is identical to the typical GC $\alpha$-enhanced RGB stars.  The
second group, which contains roughly 1/3 of the stellar population of
this GC, is very peculiar.  These stars have extremely depleted Mg,
ranging down to $-0.9$ dex below the Solar ratio, i.e., about a factor
of 15 below the normal-Mg stars.  These Mg-poor stars are identical to
those with apparently high Ca from \citeauthor*{deimos}, and from the
present detailed abundance analyses show highly enhanced K, moderately
enhanced Sc, and a small enhancement of Si and Ca compared to the
Mg-normal stars.  But there is no credible evidence for any variation
of [Fe/H] within this GC\@.  This chemical inventory is unprecedented
and unique.

We discussed whether some of this behavior, in particular the apparent
enhancement in Ca, can be attributed to low $P_e$ in the stellar
atmosphere arising from the depletion of Mg, an important electron
donor at low temperature when H is neutral.  We concluded that the
small difference in Ca abundance ($\sim$ 0.2~dex) between the Mg-poor
and Mg-normal giants is real.

A number of suggestions have been offered for for producing some of
these peculiarities, which do very rarely occur in other GC and field
halo stars, without invoking a real difference in chemical inventory
between the Mg-poor and Mg-normal giants.  We provided evidence
against the suggestion by \cite{takeda_highk} that unusually strong
turbulence in the upper atmospheres of the stars might produce the
apparent excess of K, at least in the case of NGC~2419.  
We looked at the variation with stellar
parameters and with metallicity of the non-LTE corrections for each of
the relevant elements.  In the end, we concluded that all of these
variations, correlations, and anti-correlations involving Mg, K, Sc,
Ca, and Si are real differences in mean abundances between the Mg-poor
and Mg-normal population.

It is not too difficult to imagine slightly altering the
characteristics of Type~II SNe (their mass distribution, the explosion
energy, the fallback, etc.) to reproduce the behavior of Sc and K\@.
We have not found a similar solution for Mg, Ca, and Si.  Even the
explanation for Sc and K is unsatisfactory because it requires
multiple Type~II SNe to be peculiar with respect to those SNe that
produced the material in all other known GCs.

In addition to the Mg-K anti-correlation and related issues that
we have found, there is evidence that
the usual correlations and anti-correlations among
the light elements characteristic of proton-burning at high temperature
that are seen in most GCs are present in NGC~2419 as well,
and may be accompanied by $s$-process enhancements
among some of the heavy neutron capture elements.  But the two signs
of multiple populations act independently in NGC~2419;
the Na-poor and Na-rich giants do not correspond at all with the
Mg-poor and Mg-normal giants in this peculiar GC.

With the present work, we now have a clear view of the complex
chemical inventory within NGC~2419 and of the extremely peculiar
Mg-poor population which which contains roughly 1/3 of its stellar
population.  However, we have not found a solution to the puzzle of
how to reproduce through nuclear reactions the characteristics of the
Mg-poor population in NGC~2419.  One puzzle was unveiled by
\citeauthor*{deimos}, \citeauthor*{cohen11}, and the present work, but
another has now been revealed and is at present without any
satisfactory solution.

With this new evidence demonstrating the uniqueness of NGC~2419 among
the Milky Way system of GCs, we repeat the suggestion we made in
\citeauthor*{deimos} that NGC~2419 is not a GC\@.  Instead, it may be the
nucleus of a disrupted dwarf galaxy.  Although it presently has no
dark matter \citep{baumgardt09} and a gravitational potential well
unlikely to retain supernova ejecta, it may have previously resided in
a dark matter halo, such as an accreted dwarf galaxy.  M54, the core
of the Sagittarius dwarf spheroidal galaxy, likely shares the same
origin \citep{sarajedini95}, and a similar story has been suggested
for $\omega$~Cen \citep{lee99}.  If NGC~2419 joins this growing
category of clusters, then it will be unique among its class for
retaining some (e.g., Ca, Sc, K) but not all (e.g., Fe) supernova
products.

\acknowledgements

We are grateful to the many people who have worked to make the Keck
Telescope and its instruments a reality and to operate and maintain
the Keck Observatory.  The authors wish to extend special thanks to
those of Hawaiian ancestry on whose sacred mountain we are privileged
to be guests.  Without their generous hospitality, none of the
observations presented herein would have been possible.  We thank Stan
Woosley for a helpful conversation on the nucleosynthetic origin of
potassium.  J.G.C.\ thanks NSF grant AST-0908139 for partial
support. Work by E.N.K.\ was supported by NASA through Hubble
Fellowship grant HST-HF-01233.01 awarded to E.N.K.\ by the Space
Telescope Science Institute, which is operated by the Association of
Universities for Research in Astronomy, Inc., for NASA, under contract
NAS 5-26555.

{\it Note added in proof:} A high S/N HIRES spectrum of M13 III-73 has
been obtained.  We find ${\rm [Mg/Fe]} = +0.5$~dex and ${\rm [K/Fe]} =
+0.3$~dex; both of these are normal for metal-poor GC stars.  We
cannot reproduce the unusually high K abundance claimed for this star
by \citet{takeda_highk}.

{}


\clearpage
\renewcommand{\thetable}{\arabic{table}}
\setcounter{table}{3}

\tabletypesize{\tiny}
\begin{turnpage}
\begin{deluxetable}{l r rrr rrr rrr rrr rrr rrr rrr}
\tablewidth{0pt}
\tablecolumns{23}
\tablecaption{Abundances for First 6 Red Giant Members of NGC~2419 With 
Keck/HIRES Spectra
\label{table_abund_page1} }
\tablehead{
\colhead{} & \colhead{} &
    \colhead{} & \colhead{S1131} & \colhead{} &
    \colhead{} & \colhead{S1004} & \colhead{} &
     \colhead{} & \colhead{S1065} & \colhead{} &
      \colhead{} & \colhead{S458} & \colhead{} &
       \colhead{} & \colhead{S1673} & \colhead{} &
        \colhead{} & \colhead{S1673\tablenotemark{a}} & \colhead{} &
        \colhead{} & \colhead{S1814}  & \colhead{} \\
\colhead{Elem} & \colhead{Sun} & 
    \colhead{[X/Fe]\tablenotemark{b}} & \colhead{$\sigma$\tablenotemark{c}} &
               \colhead{N} &
       \colhead{[X/Fe]} &  \colhead{$\sigma$} & \colhead{N} &
       \colhead{[X/Fe]} &  \colhead{$\sigma$} & \colhead{N} &
       \colhead{[X/Fe]} &  \colhead{$\sigma$} & \colhead{N} &
       \colhead{[X/Fe]} &  \colhead{$\sigma$} & \colhead{N} &
       \colhead{[X/Fe]} &  \colhead{$\sigma$} & \colhead{N} &
       \colhead{[X/Fe]} &  \colhead{$\sigma$} & \colhead{N} \\   
\colhead{} & \colhead{(dex)} &
     \colhead{(dex)} &  \colhead{(dex)} &  \colhead{} &
    \colhead{(dex)} &  \colhead{(dex)} &  \colhead{} &  
     \colhead{(dex)} &  \colhead{(dex)} &  \colhead{} &
     \colhead{(dex)} &  \colhead{(dex)} &  \colhead{} &
     \colhead{(dex)} &  \colhead{(dex)} &  \colhead{} &
     \colhead{(dex)} &  \colhead{(dex)} &  \colhead{} &
     \colhead{(dex)} &  \colhead{(dex)} &  \colhead{} 
}
\startdata          
C(CH)\tablenotemark{d} & 8.59 &  $-$0.73 &\nodata    &    1   &  $-$0.55 &\nodata    &    1   &  $-$0.78 &\nodata    &    1   & 
    $-$0.95 &\nodata    &    1   &  $-$1.00 &\nodata    &    1   &  $-$1.04 &\nodata    &    1   &  $-$0.77 &\nodata    &    1   \\
Na~I  &   6.32 &  $-$0.05 &   0.18    &    2   &   0.08 &   0.02    &    2   &  $-$0.02 &\nodata    &    1   &   0.10 &   0.16    &    3   &   0.44 &   0.08    &    2   &   0.39 &   0.07    &    2   &  $-$0.01 &\nodata    &    1   \\
Na~I\tablenotemark{e}      &   6.32 &\nodata &\nodata    & \nodata  &\nodata &\nodata    & \nodata  &\nodata &\nodata    & \nodata  &  
       0.04 &   0.12    &    2   &\nodata &\nodata    & \nodata  &\nodata &\nodata    & \nodata  &\nodata &\nodata    & \nodata  \\
Mg~I  &   7.54 &  $-$0.51 &   0.19    &    5   &  $-$0.16 &   0.29    &    2   &  $-$0.73 &   0.14    &    2   &  $-$0.42 &   0.14    &    4   &  $-$0.86 &   0.17    &    4   &  $-$0.88 &   0.19    &    4   &   0.60 &   0.18    &    3   \\
Mg~I\tablenotemark{f}      &   7.54 &  $-$0.54 &   0.24    &    3   &\nodata &\nodata    & \nodata  &\nodata &\nodata    & \nodata  & 
    $-$0.41 &   0.06    &    2   &  $-$1.03 &   0.09    &    2   &  $-$1.08 &   0.10    &    2   &\nodata &\nodata    & \nodata  \\
Al~I  &   6.47 &   0.43 & \nodata    &    1   &   0.70 & \nodata    &    1   &  
   $\leq$0.44 & \nodata &    1   &   0.51 &\nodata    &    1   &\nodata &\nodata    & \nodata  &\nodata &\nodata    & \nodata  & 
     0.55 &\nodata    &    1   \\
Si~I  &   7.55 &   0.59 &   0.12    &   13   &   0.42 &   0.03    &    2   &   0.50 &   0.20    &    4   &   0.61 &   0.12    &    3   &   0.79 &   0.17    &    3   &   0.67 &   0.18    &    4   &   0.43 &   0.16    &    5   \\
K~I   &   5.12 &   1.15 &\nodata    &    1   &   1.19 &\nodata    &    1   &   1.07 &\nodata    &    1   &   1.15 &\nodata    &    1   &   1.55 &\nodata    &    1   &   1.58 &\nodata    &    1   &   0.53 &\nodata    &    1   \\
\enddata
\tablenotetext{a}{This is the AGB solution for S1673.}
\tablenotetext{b}{[X/Fe]: Fe~I is used for neutral species, Fe~II for ionized species.}
\tablenotetext{c}{The rms dispersion about the mean of [X/Fe].}
\tablenotetext{d}{From synthesis of the CH band near 4320~\AA.}
\tablenotetext{e}{The strong Na D lines are omitted.}
\tablenotetext{f}{The strong Mg triplet lines are omitted.}
\tablenotetext{g}{This is [Fe(Fe~I)/H].}
\tablenotetext{h}{This is [Fe(Fe~II)/Fe(Fe~I)]].}
\end{deluxetable}

\clearpage
\end{turnpage}

\begin{turnpage}
\begin{deluxetable}{l r rrr rrr rrr rrr rrr rrr rrr}
\tabletypesize{\tiny}
\tablecolumns{23}
\tablewidth{0pt}

\tablecaption{Abundances for Last 7 Red Giant Members of NGC~2419 With 
Keck/HIRES Spectra
\label{table_abund_page2} }
\tablehead{
\colhead{} & \colhead{} &
    \colhead{} & \colhead{S1166} & \colhead{} &
    \colhead{} & \colhead{S406} & \colhead{} &
     \colhead{} & \colhead{S1209} & \colhead{} &
      \colhead{} & \colhead{S223} & \colhead{} &
       \colhead{} & \colhead{S810} & \colhead{} &
        \colhead{} & \colhead{S1305} & \colhead{} &
        \colhead{} & \colhead{S973} & \colhead{} \\
\colhead{Elem} & \colhead{Sun} & 
    \colhead{[X/Fe]\tablenotemark{a}} & 
       \colhead{$\sigma$\tablenotemark{b}} & \colhead{N} &
  \colhead{[X/Fe]\tablenotemark{a}} & \colhead{$\sigma$\tablenotemark{b}} & \colhead{N} &
  \colhead{[X/Fe]\tablenotemark{a}} & \colhead{$\sigma$\tablenotemark{b}} & \colhead{N} &
  \colhead{[X/Fe]\tablenotemark{a}} & \colhead{$\sigma$\tablenotemark{b}} & \colhead{N} &
  \colhead{[X/Fe]\tablenotemark{a}} & \colhead{$\sigma$\tablenotemark{b}} & \colhead{N} &
  \colhead{[X/Fe]\tablenotemark{a}} & \colhead{$\sigma$\tablenotemark{b}} & \colhead{N} &
  \colhead{[X/Fe]\tablenotemark{a}} & \colhead{$\sigma$\tablenotemark{b}} & \colhead{N}  \\        
\colhead{} & \colhead{(dex)} &
     \colhead{(dex)} &  \colhead{(dex)} &  \colhead{} &
    \colhead{(dex)} &  \colhead{(dex)} &  \colhead{} &  
     \colhead{(dex)} &  \colhead{(dex)} &  \colhead{} &
     \colhead{(dex)} &  \colhead{(dex)} &  \colhead{} &
     \colhead{(dex)} &  \colhead{(dex)} &  \colhead{} &
     \colhead{(dex)} &  \colhead{(dex)} &  \colhead{} &
     \colhead{(dex)} &  \colhead{(dex)} &  \colhead{} 
}
\startdata          
C(CH)\tablenotemark{c} &                         
   8.59 &  $-$0.73 & \nodata    &    1   &  $-$0.76 & \nodata    &    1   &  $-$0.74 & \nodata    &    1   &  $-$0.93 & \nodata    &    1   &  $-$0.86 & \nodata    &    1   &  $-$0.31 &   0.20    &    0   & \nodata & \nodata    &  \nodata  \\
$[$O~I$]$ &   8.83 & \nodata & \nodata    &  \nodata  & \nodata & \nodata    &  \nodata  &  
     0.70 & \nodata    &    1   & \nodata & \nodata    &  \nodata  & 
     0.24 &  0.25    &  2  & \nodata & \nodata    &  \nodata  &   0.75 &   0.04    &    2   \\
Na~I  &   6.32 &  $-$0.28 & \nodata    &    1   &   0.06 &   0.14    &    2   &   0.30 & \nodata    &    1   &   0.26 &   0.02    &    2   &  $-$0.41 &   0.15    &    3   &   0.66 &   0.15    &    2   &   0.19 &   0.15    &    2   \\
Na~I\tablenotemark{d}      &   6.32 & \nodata & \nodata    &  \nodata  & \nodata & \nodata    &  \nodata  & \nodata & \nodata    &  \nodata  & \nodata & \nodata    &  \nodata  &  $-$0.50 &   0.11    &    2   & \nodata & \nodata    &  \nodata  & \nodata & \nodata    &  \nodata  \\
Mg~I  &   7.54 &   0.42 &   0.26    &    3   &   0.36 &   0.22    &    3   &   0.44 &   0.22    &    5   &   0.14 &   0.24    &    5   &   0.24 &   0.15    &    5   &   0.26 &   0.14    &    5   &   0.29 &   0.20    &    5   \\
Mg~I\tablenotemark{e}      &   7.54 & \nodata & \nodata    &  \nodata  & \nodata & \nodata    &  \nodata  &   0.50 &   0.13    &    3   &   0.09 &   0.34    &    3   &   0.23 &   0.18    &    3   &   0.24 &   0.15    &    3   &   0.33 &   0.27    &    3   \\
Al~I  &   6.47 & \nodata & \nodata    &  \nodata  &   0.78 &   0.08    &    2   & \nodata & \nodata    &  \nodata  &   0.68 & \nodata    &    1   &   0.22 & \nodata    &    1   & \nodata & \nodata    &  \nodata  & \nodata & \nodata    &  \nodata  \\
\enddata
\tablenotetext{a}{[X/Fe]: Fe~I is used for neutral species, Fe~II for ionized species.}
\tablenotetext{b}{The rms dispersion about the mean of [X/Fe].}
\tablenotetext{c}{From synthesis of the CH band near 4320~\AA.}
\tablenotetext{d}{The strong Na D lines are omitted.}
\tablenotetext{e}{The strong Mg triplet lines are omitted.}
\tablenotetext{f}{This is [Fe(Fe~I)/H].}
\tablenotetext{g}{This is [Fe(Fe~II)/Fe(Fe~I)]].}
\end{deluxetable}

\clearpage
\end{turnpage}

\clearpage
\LongTables
\setcounter{table}{1}
\begin{deluxetable*}{lllr rrrrrr}
\tablewidth{0pt}
\tablecaption{$W_{\lambda}$ for Five Mg-Poor Stars and S406 in NGC~2419 With 
Keck/HIRES Spectra
\label{table_eqw_page1} }
\tablehead{
\colhead{Line} & \colhead{Species} & \colhead{$\chi$} & \colhead{log($gf$)} &
  \colhead{S458} & \colhead{S1004} & \colhead{S1065} & 
  \colhead{S1131} & \colhead{S1673} &  \colhead{S406} \\
\colhead{} & \colhead{} & \colhead{(eV)} &
   \colhead{(dex)} & \colhead{m\AA} & \colhead{m\AA} &
   \colhead{m\AA} & \colhead{m\AA} & \colhead{m\AA} & \colhead{m\AA}
}
\startdata
 5682.63 & Na~I  &   2.10 &  $-$0.700 &    18.6 &    17.3 &  \nodata &    13.0 &    28.6 &    15.7   \\
 5688.19 & Na~I  &   2.10 &  $-$0.420 &    23.0 &    28.1 &    27.5 &    36.0 &    53.2 &    38.5   \\
 5889.95 & Na~I  &   0.00 &   0.110 &  \nodata &   292.0 &   318.7 &  \nodata &  \nodata &   273.0   \\
 5895.92 & Na~I  &   0.00 &  $-$0.190 &   266.0 &   248.0 &   262.9 &  \nodata &  \nodata &   303.0   \\
 4703.00 & Mg~I  &   4.34 &  $-$0.440 &    73.3 &    75.4 &    48.0 &    56.0 &    28.4 &   130.2   \\
 5172.70 & Mg~I  &   2.71 &  $-$0.380 &   213.8 &   240.3 &   196.1 &   224.0 &   203.4 &   320.0   \\
 5183.62 & Mg~I  &   2.72 &  $-$0.160 &   249.0 &   275.0 &   232.5 &   266.0 &   228.0 &   390.0   \\
 5528.40 & Mg~I  &   4.34 &  $-$0.498 &    80.0 &   104.0 &    61.6 &    71.0 &    35.0 &   144.6   \\
\enddata
\end{deluxetable*}

\clearpage
\LongTables
\begin{deluxetable*}{lllr rrrrrrr}
\tablewidth{0pt}
\tablecaption{$W_{\lambda}$ for 7 Mg-Normal RGB Stars in  NGC~2419 With 
Keck/HIRES Spectra
\label{table_eqw_page2} }
\tablehead{
\colhead{Line} & \colhead{Species} & \colhead{$\chi$} & \colhead{log($gf$)} &
  \colhead{S810} & \colhead{S973} & \colhead{S1166} & 
  \colhead{S1209} & \colhead{S1305} &  \colhead{S1814} &  \colhead{S223} \\
\colhead{} & \colhead{} & \colhead{(eV)} &
   \colhead{(dex)} & \colhead{m\AA} & \colhead{m\AA} &
   \colhead{m\AA} & \colhead{m\AA} & \colhead{m\AA} & \colhead{m\AA}
   & \colhead{m\AA}
}
\startdata
 6300.30 & [O~I] &   0.00 &  $-$9.780 &  21.0 &    54.7 &  \nodata &    36.7 &  \nodata &  \nodata &  \nodata   \\
 6363.78 & [O~I] &   0.02 & $-$10.300 &  14.3 &    23.5 &  \nodata &  \nodata &  \nodata &  \nodata &  \nodata   \\
 5682.63 & Na~I  &   2.10 &  $-$0.700 &     6.0 &  \nodata &  \nodata &  \nodata &  \nodata &    30.3 &    30.3   \\
 5688.19 & Na~I  &   2.10 &  $-$0.420 &    15.0 &  \nodata &    20.3 &  \nodata &  \nodata &    46.0 &    46.0   \\
 5889.95 & Na~I  &   0.00 &   0.110 &  \nodata &   362.0 &   270.0 &  \nodata &   376.0 &  \nodata &  \nodata   \\
 5895.92 & Na~I  &   0.00 &  $-$0.190 &   236.0 &   279.0 &   304.0 &   282.0 &   355.0 &  \nodata &  \nodata   \\
 4703.00 & Mg~I  &   4.34 &  $-$0.440 &   107.0 &   123.0 &   116.5 &   125.0 &   102.3 &    92.0 &    92.0   \\
 5172.70 & Mg~I  &   2.71 &  $-$0.380 &   330.0 &   345.0 &   346.5 &   311.0 &   312.2 &   340.0 &   340.0   \\
\enddata
\end{deluxetable*}


\begin{thebibliography}{}

\bibitem[Andrievsky et al.(2010)]{mg_nonlte}
Andrievsky, S.~M., Spite, M., Korotin, S.~A.,
Spite, F., Bonifacio, P., Cayrel, R.,
Francois, P. \& Hill, V., 2010, \aap, 13223

\bibitem[Baumgardt et al.(2009)]{baumgardt09} Baumgardt, H.,
  C{\^o}t{\'e}, P., Hilker, M., et al.\ 2009, \mnras, 396, 2051

\bibitem[Baum{\"u}ller \& Gehren(1997)]{al_nonlte}
Baum{\"u}ller, D.~G. \& Gehren, T., 1997, \aap, 325, 1088

\bibitem[Briley et al.(1996)]{briley96} Briley, M.~M., Smith, V.~V.,
  Suntzeff, N.~B., et al.\ 1996, \nat, 383, 604

\bibitem[Carretta et al.(2009a)]{carretta_nao}
Carretta, E., Bragaglia, A., Gratton, R.\& Lucatello, S., 
2009, \aap, 505, 117

\bibitem[Carretta et al.(2009b)]{carretta_fe}
Carretta, E., Bragaglia, A., Gratton, R.,  D'Orazi, V. \& Lucatello, S., 
2009, \aap, 508, 695

\bibitem[Carretta et al.(2011)]{carretta_1851}
Carretta, E., Lucatello, S., Gratton, R.~G., Bragaglia, A.
  \& D'Orazi, V., 2011, \aap, 533, A69

\bibitem[Carretta et al.(2010)]{carretta_ca}
Carretta, E., Bragaglia, A., Gratton, R., Lucatello, S.,
Bellazzini, M. \& D'Orazi, V., 2010, \apjl, 712, L21

\bibitem[Carretta et al.(2012a)]{carretta_6752}
Carretta, E., Bragaglia, A., Gratton, R., Lucatello, S.
 \& D'Orazi, V., 2012, \apjl, in press (arXiv:1204.0259)
 
\bibitem[Carretta et al.(2012b)]{carretta_1851_ba}
Carretta, E., D'Orazi, V., Gratton, R.~G. \& Lucatello, S.,
2012, \aap, in press

\bibitem[Castelli \& Kurucz(2004)]{castelli04}
Castelli, F. \& Kurucz, R.~L., 2004, astro-ph/0405087

\bibitem[Cayrel et al.(2004)]{first_starsv}
Cayrel, R. et al, 2004, \aap, 416, 1117

\bibitem[Clayton(2003)]{clayton}
Clayton, D., 2003, {\it{Handbook of Isotopes in the
Cosmos}}, Cambridge University Press, Cambridge, United Kingdom

\bibitem[Cohen(1976)]{cohen_76}
Cohen, J.~G., 1976 \apjl, 203, L127

\bibitem[Cohen et al.(2004)]{emp_dwarfs}
Cohen, J.~G., Christlieb, N.,  McWilliam, A., Shectman, S.,
Thompson, I., Wasserburg, G., Ivans, I., Dehn,  Karlsson, T. \& 
Melendez, J., 2004, \apj, 612, 1107

\bibitem[Cohen \& Melendez(2005)]{cohen_m13}
Cohen, J.~G. \& Melendez, J., 2005, \aj, 129, 303

\bibitem[Cohen \& Huang(2010)]{cohen_umi}
Cohen, J.~G. \& Huang, W., 2010, \apj, 719, 931

\bibitem[Cohen et al.(2010)C10]{deimos} Cohen, J.~G., Kirby, E.~N.,
  Simon, J. \& Geha, M., 2010, \apj, 725, 288 (C10)

\bibitem[Cohen, Huang \& Kirby(2011)C11]{cohen11} Cohen, J.~G., Huang,
  W. \& Kirby, E.~N., 2011, \apj, 740, 60 (C11)

\bibitem[Cutri et al.(2003)]{2mass2}
Cutri, R.~M. et al, 2003,
``Explanatory Supplement to the 2MASS All-Sky Data Release,
http://www.ipac.caltech.edu/2mass/releases/allsky/doc/explsup.html

\bibitem[D'Antona et al.(2002)]{dantona02} D'Antona, F., Caloi, V.,
  Montalb{\'a}n, J., Ventura, P., \& Gratton, R.\ 2002, \aap, 395, 69
  
\bibitem[D'Antona et al.(2012)]{dantona12}
D'Antona, F., D'Ercole, A., Carini, R., Vesperini, E. \& Venture, P.,
\aap, in press  (arXiv:1207.1544)

\bibitem[di Criscienzo et al.(2011)]{dicriscienzo11} 
di Criscienzo, M., D'Antona, F., Milone, A.~P.,
  et al.\ 2011, \mnras, 414, 3381
  
  
\bibitem[Dupree, Hartmann \& Avrett(1984)]{dupree84} 
Dupree, A.K, Hartmann, L. \& Avrett, E.H.,  1984,  \apjl, 281, L37

\bibitem[Gratton, Sneden \& Carretta(2004)]{gratton04} Gratton, R.,
  Sneden, C. \& Carretta, E., 2004, \araa, 42, 385

\bibitem[Harris et al.(1997)]{harris97} 
Harris, W.~E., Bell, R.~A.,
  VandenBerg, D.~A., et al.\ 1997, \aj, 114, 1030
  
\bibitem[Heger \& Woosley(2010)]{heger10}
Heger, A. \& Woosley, S.~E., 2010, \apj, 724, 341   

\bibitem[Ibata et al.(1995)]{ibata95} Ibata, R.~A., Gilmore, G., \&
  Irwin, M.~J.\ 1995, \mnras, 277, 781

\bibitem[Ibata et al.(2011)]{ibata11} Ibata, R., Sollima, A., 
Nipoti, C., et al.\ 2011, \apj, 738, 186 

\bibitem[Ivanova \& Shimanski{\u i}(2000)]{ivanova_k}
Ivanova, D.~V. \& Shimanski{\u i}, V.~V., 2000, Astronomy Reports,
44, 376

\bibitem[Kirby et al.(2011)]{kirby11} Kirby, E.~N., Cohen, J.~G.,
  Smith, G.~H., et al.\ 2011, \apj, 727, 79
  
\bibitem[Kirby et al.(2010)]{kirby10} Kirby, E.~N., Guhathakurta, P.,
  Simon, J.~D., et al.\ 2010, \apjs, 191, 352
  
\bibitem[Kobayashi et al(2006)]{kobayashi06}
Kobayashi, C., Umeda, K., Nomoto K.,
  Tominaga, N. \& Ohkubo, T., 2006, \apj, 653, 1145  
  

\bibitem[Kraft(1994)]{kraft94} Kraft, R.~P.\ 1994, \pasp, 106, 553

\bibitem[Kraft et al.(1992)]{kraft92}  
Kraft, R.~P., Sneden,~C., Langer, G.~E. \& Prosser, C.~F..
 1992, \aj, 104, 645

\bibitem[Kurucz(1993)]{kurucz93} Kurucz, R. L., 1993, ATLAS9 Stellar
Atmosphere Programs and 2 km/s Grid, (Kurucz CD-ROM No. 13) 

\bibitem[Lee et al.(1999)]{lee99} Lee, Y.-W., Joo, J.-M., Sohn, Y.-J.,
  et al.\ 1999, \nat, 402, 55

\bibitem[Letarte et al.(2010)]{letarte10} Letarte, B., Hill, V.,
  Tolstoy, E., et al.\ 2010, \aap, 523, A17
  
\bibitem[Marino et al(2011)]{marino11}
Marino, A.~F. et al, 2011, \aap, 532, A8
  
\bibitem[Maeder \& Meynet(2006)]{maeder06}
Maeder, A. \& Meynet, G., 2006, \aap, 448, L37  
  
  
\bibitem[Mishenina et al.(2004)]{mishenina04} 
Mishenina, T.~V., Soubiran, C., Kovtyukh, V.~V., \& Korotin, S.~A.\ 2004, 
\aap, 418, 551
  
\bibitem[Mucciarelli et al.(2012)]{mucciarelli} Mucciarelli, A.,
  Bellazzini, M., Ibata, R., Merle, T.  \& Chapman, S.~C., 2012,
  \mnras, submitted

\bibitem[Nomoto et al.(2006)]{nomoto06} Nomoto, K., Tominaga, N.,
  Umeda, H., Kobayashi, C., \& Maeda, K.\ 2006, Nuclear Physics A,
  777, 424


\bibitem[Norris et al.(1996)]{norris96}
Norris, J.~E., Freeman, K.~C. \& Mighell, K.~J., 1996, \apj, 462, 241
  
    
\bibitem[Pilachowski et al.(1996)]{pilachowski96}
Pilachowski, C.~A.,  Sneden, C.,  Kraft, R.~P. \& Langer, G.~E., 
1996, \aj, 112, 545

\bibitem[Puzia, Perrett \& Bridges(2005)]{puzia05}
Puzia, T.~H., Perrett, K.~M. \& Bridges, T.~J., 2005, \aap, 434, 090

\bibitem[Sarajedini \& Layden(1995)]{sarajedini95} Sarajedini, A., \&
  Layden, A.~C.\ 1995, \aj, 109, 1086

\bibitem[Shetrone, C\^ot\'e \& Sargent(2001)]{shetrone01}
Shetrone,~M., C\^ot\'e, P. \& Sargent, W.~L.~W., 2001, \apj, 568, 592  

\bibitem[Shimanskaya et al.(2000)]{shimanskaya00} Shimanskaya, N.~N.,
  Mashonkina, L.~I., \& Sakhibullin, N.~A.\ 2000, Astronomy Reports,
  44, 530 

\bibitem[Skrutskie et al.(2006)]{2mass1}
Skrutskie, M.~F. et al, 2006, \aj, 131, 1163

\bibitem[Sneden(1973)]{moog} Sneden, C., 1973, Ph.D. thesis, Univ.
of Texas

\bibitem[Sobeck et al.(2011)]{sobeck_moog}
Sobeck, J.~S. et al, 2011, \aj, 141, 175

\bibitem[Spite et al.(2012)]{spite2012}
Spite, M., Andrievsky, S~M., Spite, F. et al, 2012,
\aap, in press (arXiv:1204.1139)

\bibitem[Stetson(2005)]{stetson05}
Stetson, P.~B., 2005, \pasp, 117, 563

\bibitem[Suntzeff et al.(1988)]{suntzeff88} Suntzeff, N.~B., Kraft,
  R.~P., \& Kinman, T.~D.\ 1988, \aj, 95, 91
  

  
\bibitem[Takeda et al.(2002)]{takeda_knonlte}
Takeda, Y., Zhao, G., Chen, Y.~Q., Qui, H.~M.
\& Takada-Hidai, M., 2002, PASJ,54,275 

\bibitem[Takeda et al.(2010)]{takeda_highk}
Takeda, Y., Kaneko, H., Matsumoto, N., Oshino, S.,
Ito, H. \& Shibuya, T.,  2010, PASJ, 61, 563

\bibitem[Tominaga, Umeda \& Nomoto(2007)]{tominaga07}
Tominaga, N., Umeda, H. \& Nomoto, K., 2007, \apj, 660, 516 

\bibitem[VandenBerg et al.(2012)]{vandenberg12} VandenBerg, D.~A.,
  Bergbusch, P.~A., Dotter, A., Ferguson, J., Michaud, G., Richer,
  J. \& Profitt, C.~R., 2012,, \apj, in press (arXiv:126.1820)
      
\bibitem[Vogt et al.(1994)]{vogt_hires} 
Vogt, S.~E. et al.\, 1994, SPIE, 2198, 362

\bibitem[Worthey \& Ottaviani(1997)]{worthey97}
Worthey, G. \& Ottaviani, D.~L., 1997, \apjs, 111, 377


\bibitem[Yi et al.(2003)]{yi03}
Yi, S.,  Kim, Y.-C., Demarque, P. \& Alexander, D.~R., 2003, \apjs, 143, 499

\bibitem[Yong et al.(2005)]{yong_6752}
Yong, D., Grundahl, F., Nissen, P.~E., Jensen, H.~R.
\& Lambert, D.~L., 2005, 2005, \aap, 438, 875

\bibitem[Yong \& Grundahl(2008)]{yong08}
Yong, D. \& Grundahl, F., 2008, \apj, 672, L29


\end{thebibliography}
\end{document}